\documentclass[journal=jacsat,manuscript=article]{achemso}
\setkeys{acs}{articletitle = False}
\SectionNumbersOn

\usepackage{amsmath}
\usepackage{multirow}
\usepackage{xspace}
\usepackage{booktabs}
\usepackage{float}
\usepackage{multirow}
\usepackage{subfig}
\usepackage{xcolor}

\newcommand*{\kcal}{kcal~mol$^{\textrm{-}1}$\xspace}

\newcommand*{\degree}{\ensuremath{^\circ}\xspace}
\newcommand*{\abinitio}{{\it ab initio}\xspace}
\newcommand*{\etal}{{\it et al.}\xspace}

\author{Yudong Qiu}
\affiliation[UCD]{Chemistry Department\\
The University of California, Davis\\
Davis, California 95616}
\author{Benedict R. Schwegler}
\affiliation[STAN]{Department of Civil and Environmental Engineering\\
Stanford University\\
Stanford, CA 94305}
\author{Lee-Ping Wang}
\affiliation[UCD]{Chemistry Department\\
The University of California, Davis\\
Davis, California 95616}
\email{leeping@ucdavis.edu}

\title[Membrane]{Polarizable Molecular Simulations Reveal How Silicon-containing Functional Groups Govern the Desalination Mechanism in Nanoporous Graphene}

\begin{document}

\begin{abstract}
We report a molecular dynamics (MD) simulation study of reverse osmosis desalination using nanoporous monolayer graphene passivated by SiH$_2$ and Si(OH)$_2$ functional groups.
A highly accurate and detailed polarizable molecular mechanics force field model was developed for simulating graphene nanopores of various sizes and geometries.
The simulated water fluxes and ion rejection percentages are explained using detailed atomistic mechanisms derived from analysis of the simulation trajectories. Our main findings are:
(1) The Si(OH)$_2$ pores possess superior ion rejection rates due to selective electrostatic repulsion of Cl$^-$ ions, but Na$^+$ ions are attracted to the pore and block water transfer.
(2) By contrast, the SiH$_2$ pores operate via a steric mechanism that excludes ions based on the size and flexibility of their hydration layers.
(3) In absence of ions, water flux is directly proportional to the solvent accessible area within the pore; however, simulated fluxes are lower than those inferred from recent experimental work.
We also provide some hypotheses that could resolve the differences between simulation and experiment.
\end{abstract}

\newpage

\section{Introduction}

Seawater desalination plays a crucial role in the research for overcoming water scarcity, one of the most serious challenges humans are facing.\cite{Shannon:2008bk, Elimelech:2011bs}
Multiple approaches have been investigated, including multi-stage flash,\cite{VANDERBRUGGEN2002207} membrane distillation,\cite{VANDERBRUGGEN2002207} mechanical vapor compression,\cite{ALJUWAYHEL1997253} and reverse osmosis (RO).\cite{Greenlee:2009dt}
Among them, reverse osmosis technology is by far the most promising method, as it requires the least amount of energy input to purify the same amount of water.
Conventional RO membranes made of composite materials, like polymers, has improved over decades to reach an acceptable energy efficiency;\cite{Busch:2004dx}
this has led the growth of RO plants all over the world, but improvements in membrane permeability have been limited since the 1990s.\cite{LEE20111}
Therefore, people seek new forms of materials, such as carbon nanotubes,\cite{Thomas:2008bc} and nanoporous graphene\cite{CohenTanugi:2012hs,Surwade:2015cy} that hold promise for desalination applications.

Single-layer nanoporous graphene is a promising desalination membrane due to its unmatched thinness, good mechanical strength, and chemical stability.
Experimental studies on nanoporous graphene involve various treatments to create nanopores in graphene, including electrical pulse,\cite{Rollings:2016hz} ion bombardment,\cite{OHern:2014cv} and O$_2$ plasma etching.\cite{Surwade:2015cy}
Ion passage through the prepared nanoporous graphene was observed to be highly selective in several studies\cite{OHern:2014cv, Rollings:2016hz, Hong:2017cn}, with cation/anion selectivity ratio of over 100; ion selectivity is positively correlated with salt rejection.\cite{Rollings:2016hz}
Recently, one experimental study has shown nanoporous graphene can act as a desalination membrane with surprisingly high water flux and nearly perfect salt rejection.\cite{Surwade:2015cy}
The reported 10$^6$ g m$^{-2}$ s$^{-1}$ water flux (3000 water molecules per pore per ps) under only 0.17 bar driving pressure is several orders of magnitude higher than conventional membranes characterized by fluxes of about 12 g m$^{-2}$ s$^{-1}$ under 83 bar.\cite{Busch:2004dx}
Scanning transmission electron microscopy (STEM) images suggested that the pores produced by O$_2$ plasma were around 1 nm in diameter.
Lee \etal \cite{Lee:2014gs} proposed that graphene nanopores could be stabilized by silicon based on STEM imaging and density functional calculations; these results suggest that the edges of nanopores produced by oxygen plasma may also contain silicon.
These studies have contributed essential insights into the nanopore structure; however, the molecular mechanism of water transport and ion rejection responsible for the high performance remain unknown,
creating an opportunity for theory and simulation to provide detailed insights into the desalination process in nanoporous graphene.

Several published theoretical studies have focused on the possibility of nanoporous graphene as a desalination membrane.
In 2012, Cohen-Tanugi and Grossman investigated the use of hydrogen- or hydroxyl- passivated graphene nanopores to separate NaCl ions from water.\cite{CohenTanugi:2012hs}
Their molecular dynamics (MD) study suggested a maximum diameter of 5.5 \AA\ for maintaining a high percentage of salt rejection.
They also followed up the studies with reduced graphene oxide\cite{Lin:2015jy} and multilayer porous graphene.\cite{CohenTanugi:2016en}
Strong and Eaves investigated the hydrodynamics of the water transferring through porous graphene\cite{Strong:2016iua}
and suggested that the non-equilibrium dynamics of the water transfer is controlled by two competing mechanisms---translocation mechanism and evaporation-condensation mechanism.
Ebrahimi\cite{EBRAHIMI2016160} modeled the nanoporous graphene passivated by Si atoms and found that the membrane curvature reduces the water flux.
The relation between pore ion selectivity and surface charge was studied by Zhao \etal\cite{Zhao:2013ku}, who predicted that negative charges at pore edges could impede the passage of Cl$^-$.
Very recently, Li \etal \cite{Li:2018eu} found that the ion selectivity of multi-layer graphene nanopores may decrease if the surface charge becomes too strong due to counter-ions entering the pore and causing a charge inversion.
These studies all used fixed-charge models that do not account for the environmental dependence of molecular dipole moments, raising the possibility that a more detailed physical model is needed to explain the water transfer phenomenon at the interface of the nanopore.

The inclusion of polarizability greatly improves on the physical descriptions of molecular mechanics (MM) force field models by explicitly modeling induced polarization from external electric fields or intermolecular interactions.\cite{Baker:2015fa}
In particular, the AMOEBA model includes polarizable point dipoles and also incorporates atomic multipole moments up through the quadrupole,\cite{Ren:2003dn,Ponder:2010jva,Ren:2011bqba,Laury:2015da} providing a more detailed description of molecular electrostatics compared to other widely implemented polarizable force fields such as the Drude oscillator\cite{Lamoureux:2003cb,Lopes:2013fn} and the fluctuating-charge models\cite{Chen:2008bs, Bauer:2012kd}.
As a result, AMOEBA models can reach quantitative agreement against gas-phase \abinitio interaction energies at various distances and orientations, and produce reliable predictions when the models are used to simulate condensed-phase phenomena.\cite{Zhang2012,Yue2013,Mu2014,Yue2015,Bell2016}
Because molecular polarization depends on changes in the electrostatic environment, we expect a polarizable model to yield a more accurate description of water transfer and ion rejection as water molecules and ions move from the bulk and through the nanopore.

In this paper we describe an accurate ``AMOEBA-type'' polarizable force field for silicon-passivated nanoporous graphene, and its application in MD simulations on model nanopores that reveal their performance in desalination.
We find that hydrophobic nanopores, passivated by SiH$_2$ groups (also denoted as Si--H), has superior water transfer rates compared to hydrophilic Si(OH)$_2$ nanopores (also denoted as Si--OH) ; on the other hand, hydrophilic Si--OH nanopores selectively block the negatively charged Cl$^-$ ions resulting in an improved salt rejection.
We also provide new atomistic insights into the physical mechanisms that govern water transfer rates and salt rejection percentages on the atomic scale; one of our novel observations is that ions attracted by hydrophilic functional groups may block the pores, thereby significantly reducing water transfer rates.
However, under the same assumed pore density as in the experimental work, our calculated transfer rates are still two orders of magnitude lower than experiment\cite{Surwade:2015cy}, suggesting there still exist differences between the experimental result and our molecular picture.


The rest of the paper is organized as follows.
First, we describe the development of our polarizable nanopore model and force field, the setup of desalination simulations, and the simulation analysis methodology.
Next, we describe the optimized parameters and simulation results, including our simulated water transfer and ion rejection rates and discussions of the molecular mechanisms.
Finally, we provide some hypotheses that may explain the remaining differences between simulation and experiment.

\newpage
\section{Methods and Computational Details}

\subsection{Polarizable Model}

The general functional form of the AMOEBA polarizable force field can be divided into four terms:

\begin{equation}
  \label{energy_terms}
  U = U_\mathrm{bonded} + U_\mathrm{VdW} + U_\mathrm{ele}^\mathrm{perm} + U_\mathrm{ele}^\mathrm{ind}
\end{equation}

The first term describe valence interactions which include bond stretching, angle bending, and torsional rotations; they may also include out-of-plane bending terms and higher-order couplings.
The last three terms describe the non-bonded interactions, including the van der Waals (vdW) force and the electrostatic contributions from permanent and induced multipoles.
Two types of local molecular frames used for defining the permanent dipole and quadrupole moments are shown in the Supporting Information. (Figure~S1 and Table~S6).

\begin{figure}[H]
  \centering
  \includegraphics[width=0.8\textwidth]{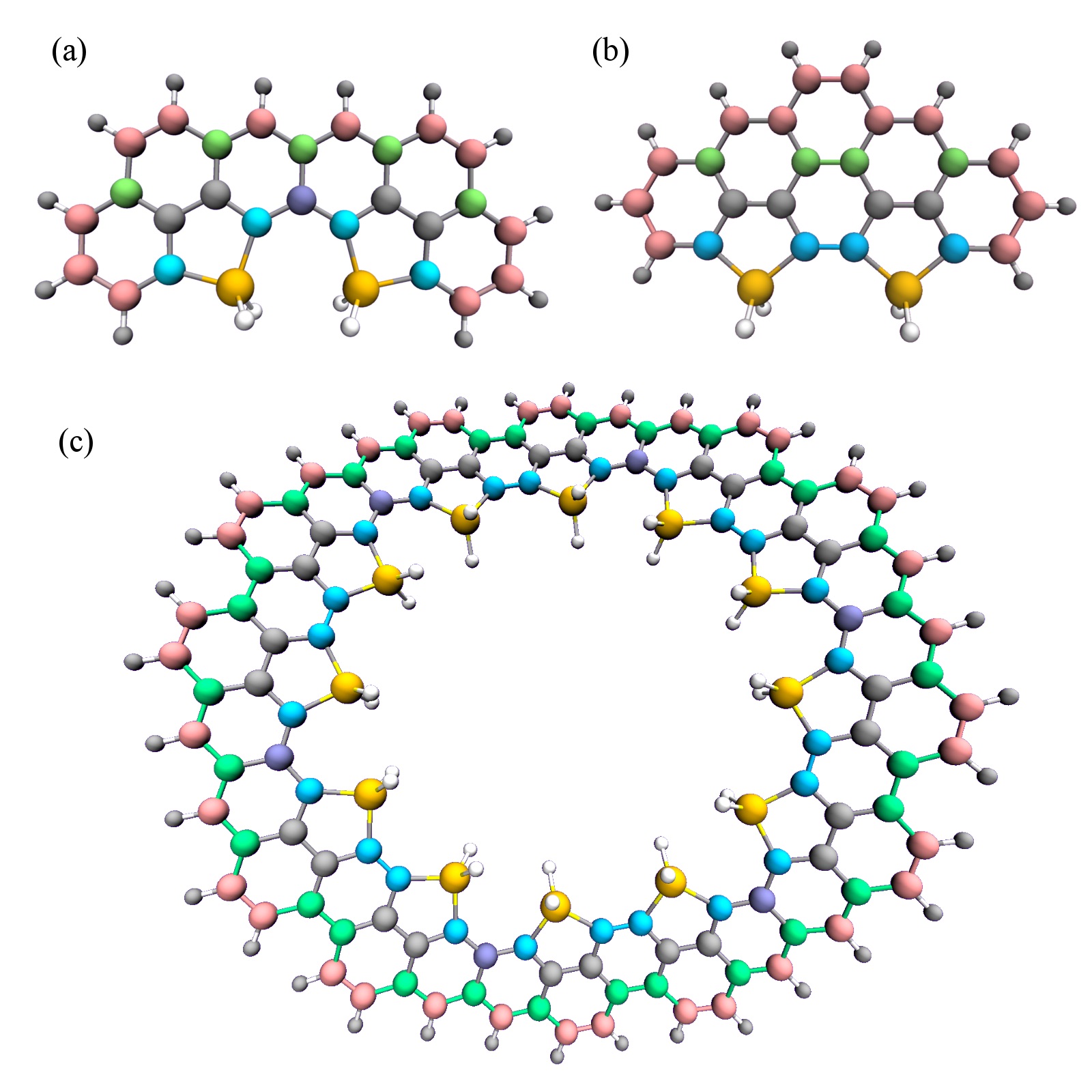}
  \caption{Parameterization models. (a) MED: Medium-sized model  with Si-C-C-C-Si structure. (b) MED2: Medium-sized model with Si-C-C-Si structure. (c) LRG: Large model containing an entire ring of 12 Si atoms. Three elements included are silicon (large spheres), carbon (medium), hydrogen (small). Carbon atom types: pink--Ca1, green--Ca2, grey--Cb, purple--Cc, blue--Cs; Silicon: yellow--Si; Hydrogen: white--Hs, grey--Ha. }
  \label{fb_models}
\end{figure}

We constructed an atomistic model of nanoporous graphene passivated by Si atoms in the form of five-membered rings according to previous STEM imaging and density functional studies.\cite{Lee:2014gs}
Figure~\ref{fb_models} shows the molecular models used in QM calculations to develop of force field parameters for the Si--H pore.
We constructed two medium-sized molecules, C$_{26}$H$_{16}$Si$_2$ (Figure~\ref{fb_models}a) and C$_{24}$H$_{14}$Si$_2$ (Figure~\ref{fb_models}b), to characterize both the Si--C--C--C--Si and the Si--C--C--Si structures in the complete pores.
We also created a small model incorporating only one Si atom (SI Figure~S2), but it was not included in the final data set because it lacks the interaction between Si passivating groups.
For carbon atoms, five different atom types, namely Ca1, Ca2, Cb, Cc, Cs, were created based on the topology and their distance to the pore center, as shown in different colors. Figure~\ref{fb_models}c illustrates how a complete Si$_{12}$H$_{24}$ pore can be built by assembling the two types of medium-sized models. This large model contains 120 carbon, 12 silicon, and 60 hydrogen atoms.

\begin{figure}[H]
    \centering
      \includegraphics[width=0.8\textwidth]{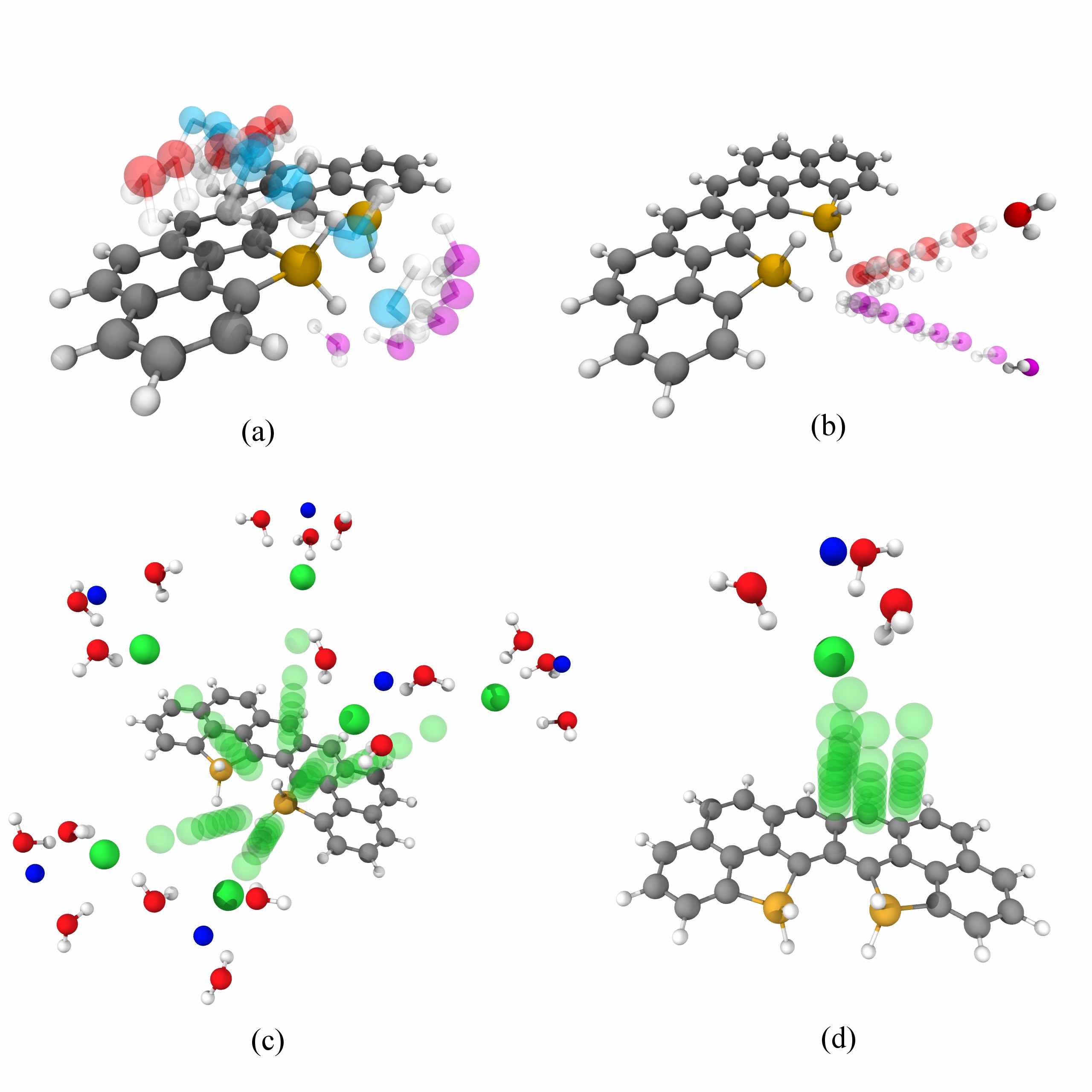}
    \caption{Sample structures for computing pore--water and pore--ion interaction energies. (a) Scanning the water molecule over the medium model. Three colors indicate the scan in three directions: Red--X, Blue--Y, Purple--Z. (b) Varying water distances and orientations. Two colors of water show two orientations. (c) Samples with various Si--Cl distances. (d) Samples with various C--Cl distances. Colors: Silver--C, White--H, Yellow--Si, Red--O, Green--Cl$^-$, Blue--Na$^+$. In C and D each transparent Cl$^-$ ion represents a complete [Na$^+$(H$_2$O)$_3$Cl$^-$] group, and the [Na$^+$(H$_2$O)$_3$] fragment follows the Cl$^-$ ion.}
  \label{interaction_water_ion}
\end{figure}

Table~\ref{ref_data} summarizes the reference data set for the Si--H pore model.
In order to generate data describing the intramolecular forces, we carried out \abinitio MD simulations at the B3LYP/6-31G* level of theory\cite{PhysRev.140.A1133} using the TeraChem software package\cite{Ufimtsev:2009em,Titov:2013iw}.
Under the simulated temperature of 1500 K, 300 snapshots were generated with an 0.5 fs time interval, from which single-point energies and gradients were computed at the DF-MP2/cc-pVTZ level of theory using the Psi4 software package\cite{Parrish:2017hg}.
We then characterized the non-bonded interactions by computing interaction energies at various model geometries including some ``hand-constructed'' structures and results of constrained optimizations.
Figure~\ref{interaction_water_ion}a gives an example of such a calculation where we scanned the position of the water molecule parallel to the molecular plane of the medium pore model using the translation-rotation-internal coordinate (TRIC) geometry optimizer;\cite{Wang:2016ed} these scans were performed in three directions (X, Y, \& Z) as indicated in Figure~\ref{interaction_water_ion}a.
For several selected geometries, we also varied the distance between the water oxygen atom and pore H atom in Si--H group, as well as the water molecular orientation. (Figure~\ref{interaction_water_ion}b)
To characterize the interaction between pore and ions, we built a partially solvated [Na(H$_2$O)$_3$Cl] cluster model and positioned it in two linear orientations such that Si--Cl--Na or Si--Na--Cl are on a straight line.
Sample configurations were generated by constrained optimizations where we scanned the distance between various pore model atoms and the ion directly facing the pore (Figure~\ref{interaction_water_ion}c and d).
For the geometries collected above, we computed counterpoise-corrected binding energies at DLPNO-CCSD(T) level of theory with the def2-TZVP basis set\cite{Neese:2009db,Riplinger:2013dx} implemented in ORCA\cite{Neese:2012vd}.
This approximate high-level \abinitio method allows us to obtain accurate energetics for relatively large molecules; our preliminary benchmarks show that the binding energy of a smaller pore--water model C$_{8}$H$_{8}$Si computed by DLPNO-CCSD(T) method has a difference less than 0.1 \kcal compared to the original CCSD(T) method. (SI section 2)
The reference data set consists over 1100 quantum interaction energies for the Si--H MED/MED2 pore models.
The water--``large pore" model contains 195 atoms and 4567 basis functions with def2-TZVP basis set, which limited the number of calculations that we could afford to only four.
We included them mainly as a sanity check to ensure the results did not deviate significantly from the medium models.
The entire procedure was repeated for the Si--OH pore models, in which all of the SiH$_2$ terminating groups were replaced by Si(OH)$_2$ groups.

\begin{table}[H]
\begin{flushleft}
\caption{Description of Reference Data for Parameterization of Si--H Pore Model.}
\label{ref_data}
{
\renewcommand{\tabcolsep}{5.5pt}
\begin{tabular}{l|l|l|l}
\toprule
\multicolumn{2}{c|}{Reference Data}                                                                                       &  Method              &  No. Calcs.              \\ \hline
\multicolumn{2}{l|}{MED Potential Energies and Forces}                                                                    &  A                   &  300                     \\ \hline
\multirow{3}{*}{\begin{tabular}{l}MED--Water \\Interaction Energies\end{tabular}}      &  Constrained Optimize X, Y, Z    &  B                   &  163                     \\
                                                                                       &  Energies v.s. Distances for     &  \multirow{2}{*}{B}  &  \multirow{2}{*}{260}    \\
                                                                                       &  X, Z Scans (2 orientations)     &                      &                          \\ \hline
\multirow{2}{*}{\begin{tabular}{l}MED--Ion \\Interaction Energies\end{tabular}}        &  Energies v.s. Distances for     &  \multirow{2}{*}{B}  &  \multirow{2}{*}{180}    \\
                                                                                       &  Si--Ion and C--Ion              &                      &                          \\ \hline
\multicolumn{2}{l|}{MED2 Potential Energies and Forces}                                                                   &  A                   &  300                     \\ \hline
\multirow{3}{*}{\begin{tabular}{l}MED2--Water \\Interaction Energies\end{tabular}}     &  Constrained Optimize X, Y, Z    &  B                   &  163                     \\
                                                                                       &  Energies v.s. Distances for     &  \multirow{2}{*}{B}  &  \multirow{2}{*}{175}    \\
                                                                                       &  X, Z Scans (2 orientations)     &                      &                          \\ \hline
\multirow{2}{*}{\begin{tabular}{l}MED2--Ion \\Interaction Energies\end{tabular}}       &  Energies v.s. Distances for     &  \multirow{2}{*}{B}  &  \multirow{2}{*}{162}    \\
                                                                                       &  Si--Ion and C--Ion              &                      &                          \\ \hline
\multicolumn{2}{l|}{Coronene Potential Energies and Forces}                                                               &  A                   &  300                     \\ \hline
\multirow{2}{*}{\begin{tabular}{l}Coronene--Water \\Interaction Energies\end{tabular}} & Energies v.s. Distances          &  \multirow{2}{*}{B}  &  \multirow{2}{*}{240}    \\
                                                                                       & with different orientations      &                      &                          \\ \hline
{\begin{tabular}{l}LRG--Water \\Interaction Energies\end{tabular}}                     & Energies at optimized geometries &  B                   &  4                       \\ \hline
\multicolumn{2}{l|}{Total}                                                                                                &                      &  2247                    \\
\bottomrule
\end{tabular} \\
A. DF-MP2/cc-pVTZ on geometries from B3LYP/6-31G* AIMD simulations. \\
B. DLPNO-CCSD(T)/def2-TZVP on geometries optimized at B3LYP/6-311G* level of theory.
}
\end{flushleft}
\end{table}

The graphene carbon atoms more than four bonds away from Si atoms were described using one extra atom type, Ca0, which was not included in these models. 
The interaction parameters of Ca0 were fitted to coronene--water binding energies  (Supporting Information Section S3).
Although the modeling of graphene with finite polycyclic aromatic hydrocarbons is imperfect due to the closing of the HOMO-LUMO gap and emerging radical character at the graphene edges\cite{Pykal:2016eu}, Lazar \etal have shown that for seven small organic molecules, experimental adsorption enthalpies on graphene can be well predicted by \abinitio adsorption enthalpies on coronene.\cite{Lazar:2013jm}
To ensure transferability from the finite-sized coronene model systems to the continuous graphene model, we fixed the charge of Ha atom type (C-H) and both charge and dipole parameters for the Ca0 atom type to be zero.

The total charge of the large model was balanced to zero in our parameterization.
In keeping with our assumption that the nanoporous graphene should be overall neutral in the simulations, the simulation setup of various pores included a uniform neutralizing charge to all ``pore'' atoms with a magnitude that never exceeded 0.0001 elementary charge.

Following the generation and organization of \abinitio data, the force field parameters were optimized using ForceBalance\cite{Wang:2014gy}.
An initial set of bonded parameters and vdW parameters were adopted from the carbon and hydrogen parameters of benzene and methane in the AMOEBA09 force field for organic molecules.\cite{Ren:2011bqba}
For the electrostatic interactions, we followed the protocol utilizing the GDMA program,\cite{Stone:2005hs} described in the supporting information of the AMOEBA09 paper\cite{Ren:2011bqba} to generate the initial guess for the charge, dipole, and quadrupole moments.
The AMOEBA03 water and ion model parameters were employed in our parameterization and simulation.\cite{Grossfield:2003ep,Ren:2003dn}

\subsection{Simulation Setup}

\begin{figure}[H]
    \centering
    \includegraphics[width=0.8\textwidth]{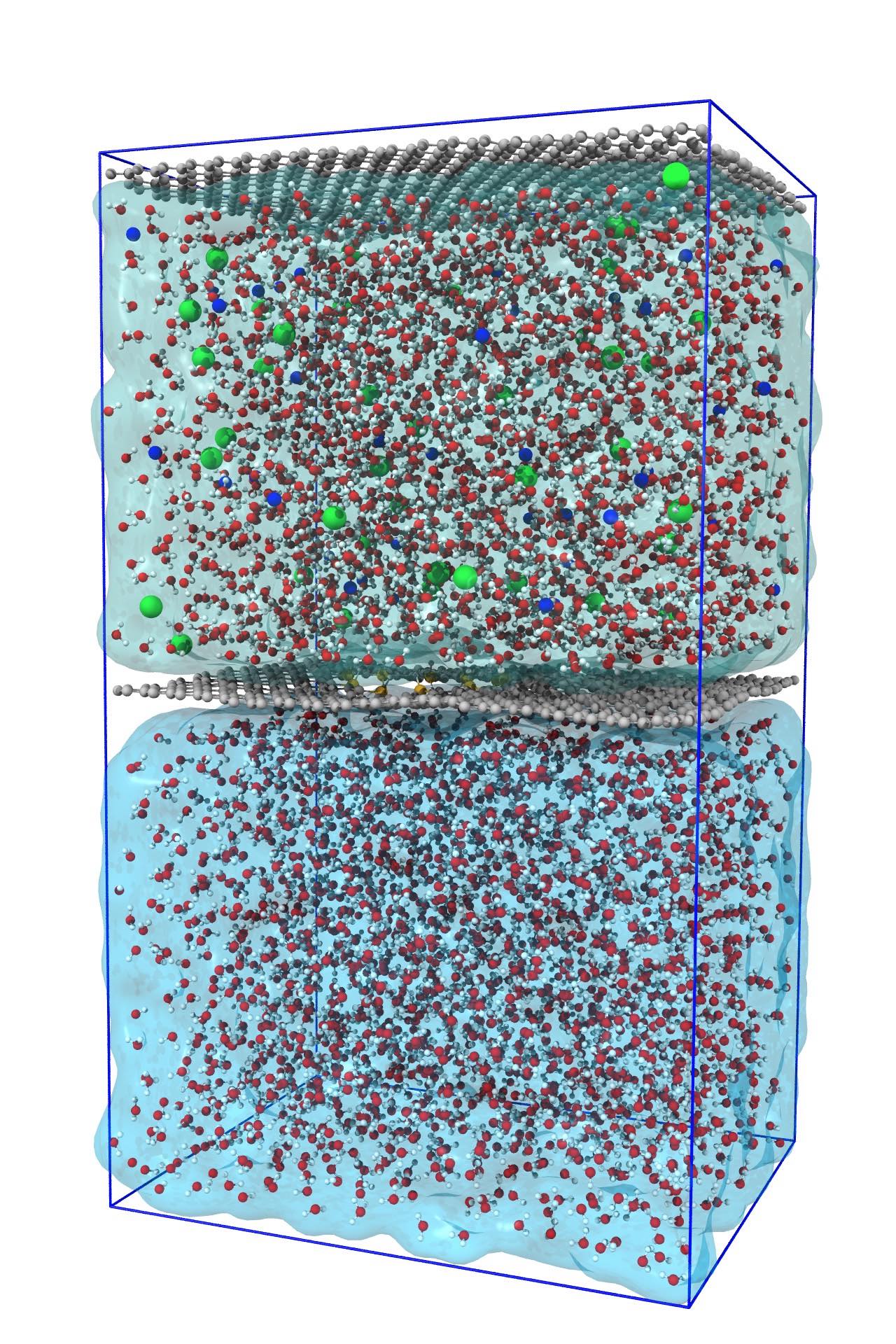}
    \caption{Simulated system in 3D periodic boundary conditions. Top half represents the feed side (green surface) and bottom half represents permeate side (blue surface). Particles: Green--Cl$^-$, Blue--Na$^+$, Grey--C, Yellow--Si, Red--O, White--H.}
    \label{simulation_box}
\end{figure}

Figure~\ref{simulation_box} depicts the simulated sytem. The simulation box is 5 nm $\times$ 4 nm $\times$ 8 nm with 3D periodic boundary conditions, and contains around 4000 water molecules together with 40 Na$^+$ and 40 Cl$^-$ ions corresponding to 1 M NaCl solution. The simulation box is partitioned into the feed and permeate side by the nanoporous graphene centered on the z-axis of the simulation cell and a second graphene sheet without pores at the top edge.

\begin{figure}[H]
    \centering
    \includegraphics[width=0.8\textwidth]{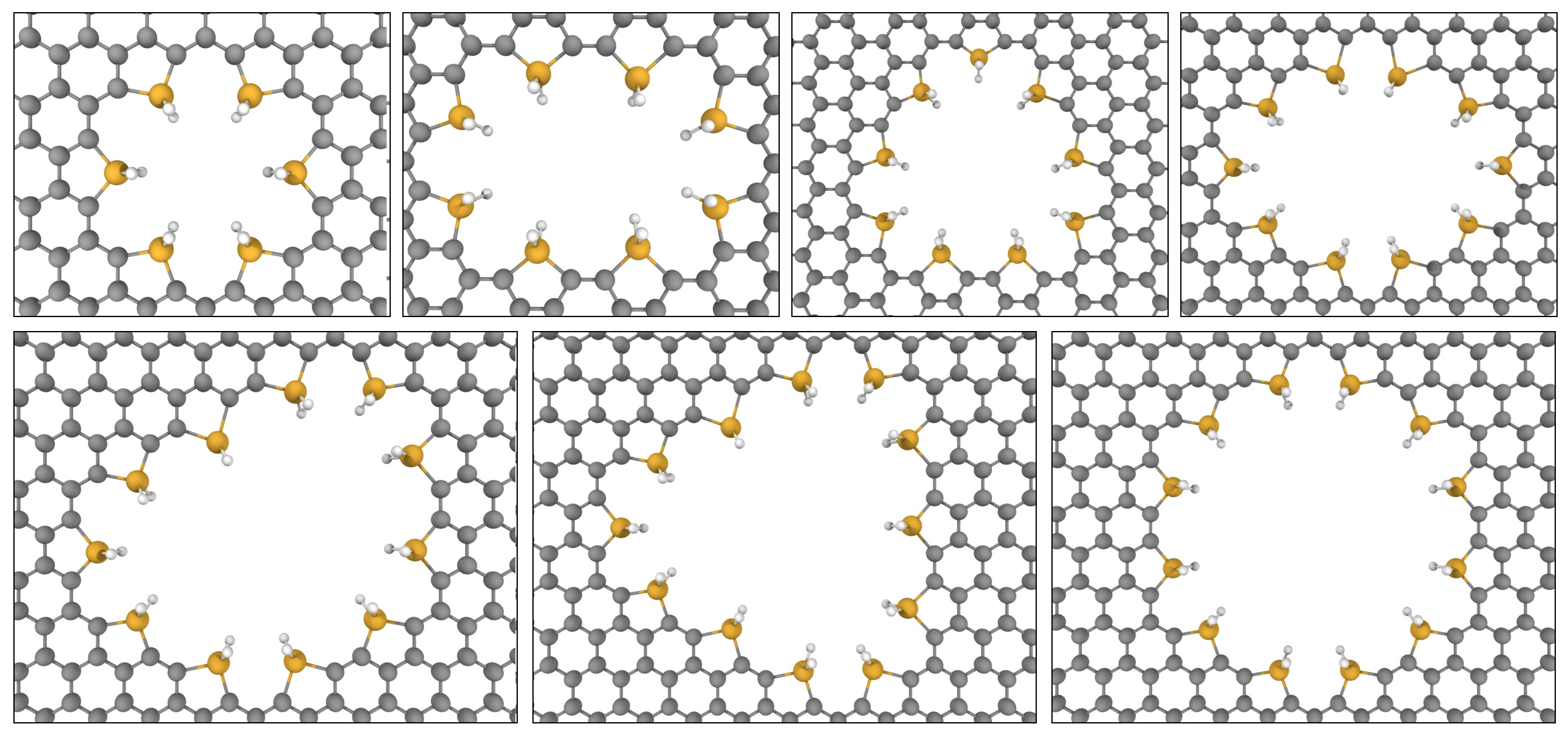}
    \caption{Silicon-passivated nanopore models. Top row: Si-6, Si-8, Si-9, Si-10. Bottom row: Si-11, Si-12, Si-12s.}
    \label{pore_models}
\end{figure}

Seven pore structures of different sizes were constructed as shown in Figure~\ref{pore_models}. According to the number of Si atoms on the edge, they are labeled as Si-6, Si-8, Si-9, Si-10, Si-11, Si-12, Si-12s respectively. The Si--OH pores have analogous structures where all Si--H groups are replaced by Si--OH groups.

All desalination simulations used a Langevin thermostat with a collision frequency of 0.1 ps$^{-1}$ to maintain a temperature of 348.15 K.
The particle-mesh Ewald (PME) method\cite{Ren:2003dn} with cut-off distance at 0.9 nm was adopted when evaluating non-bonded interactions.
The systems are first equilibrated for 5 ns with step length of 1 fs; a Monte Carlo barostat with 1 atm pressure was used to adjust the size of the periodic box once every 25 MD steps.
All three dimensions of the periodic box were allowed to change individually to relax internal tensions within the graphene model.
To prevent ions from leaking during this equilibration period, a virtual blocking force is added that occupies the space of the pore and only repels the ions (Supporting Information Section S5).

Following equilibration, the volume is held constant and an external force is applied to the graphene wall to simulate the driving pressure, pushing water from the feed side to the permeate side.
Simulation data is collected for 10 ns and repeated ten times from different equilibrated geometries to obtain averages and standard error estimates, for a total of 100 ns per system.
Each of the seven Si--H and seven Si--OH pores were simulated at four driving pressures (100, 200, 500 and 1000 bar) for a total of 5.6 $\mu$s of simulation data in the main study.
We also performed several simulations with the graphene removed and/or ions removed to address specific questions, described in later sections.

\subsection{Trajectory Analysis}

\begin{figure}[H]
  \centering
    \subfloat[][]{
    	\includegraphics[width=.49\linewidth]{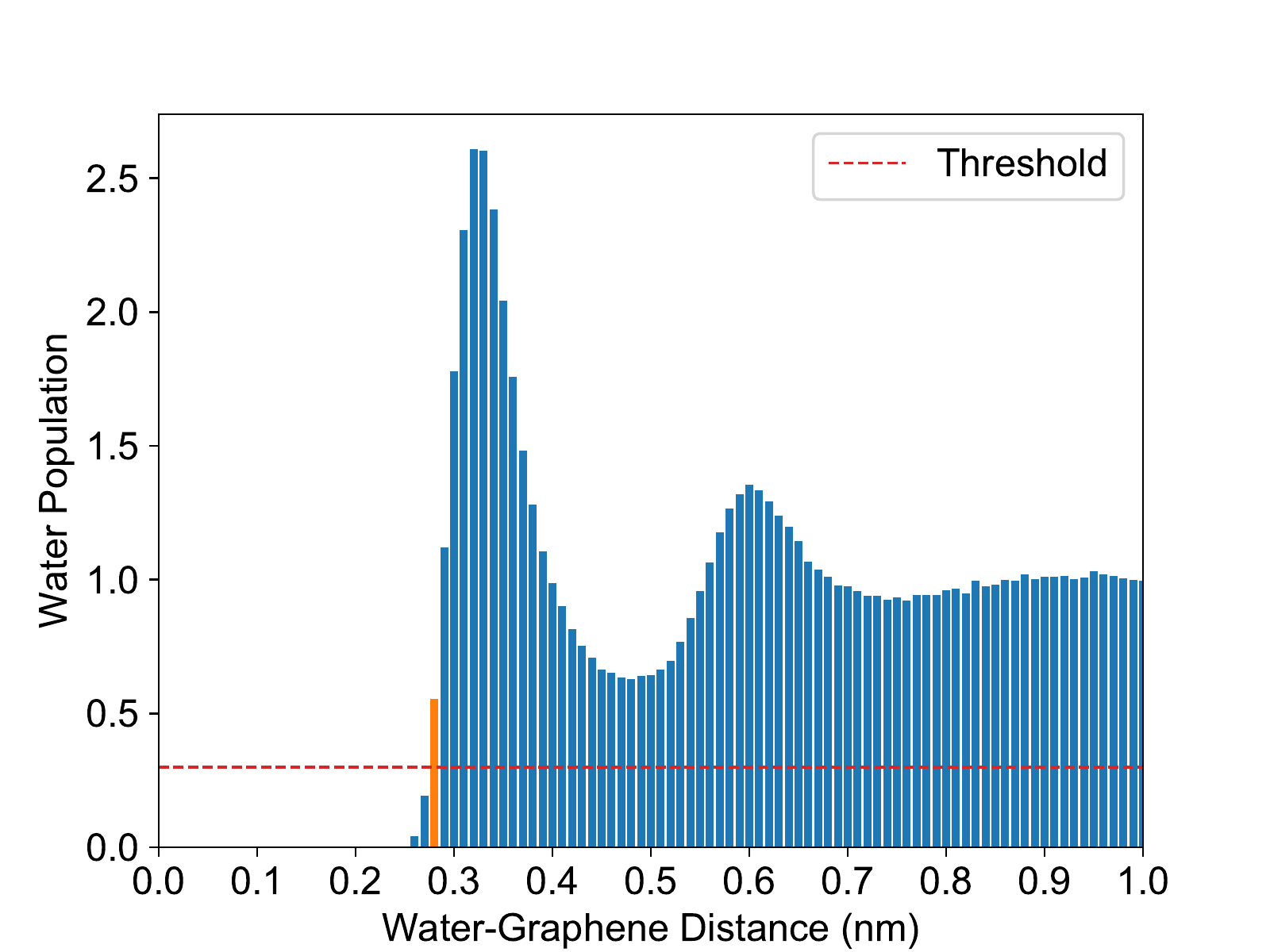}
	\label{thickness}
	}
    \subfloat[][]{
    	\includegraphics[width=.49\linewidth]{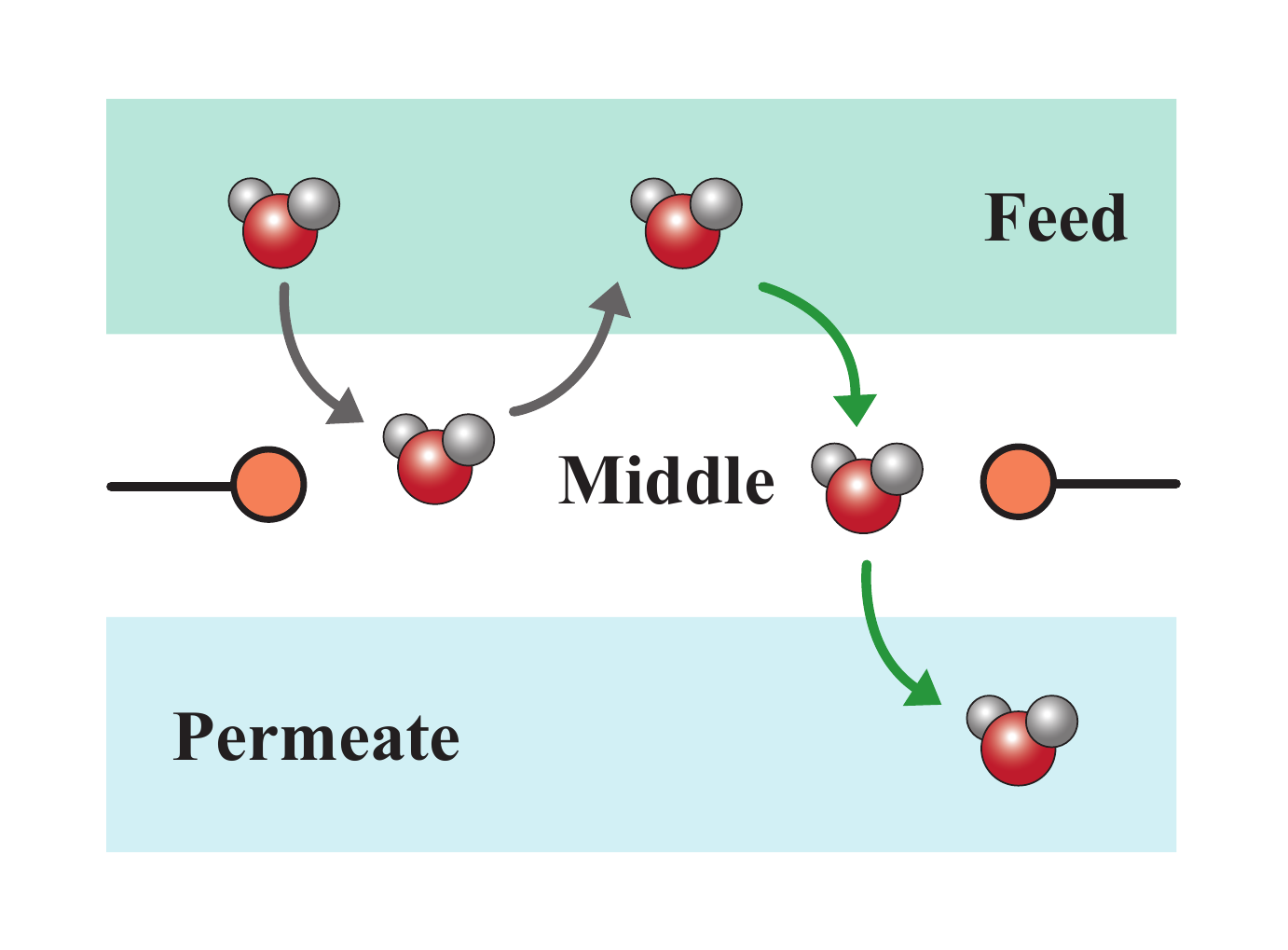}
	\label{transfer_diagram}
	}
  \caption{\protect\subref{thickness} Radial distribution histogram between water and nanoporous graphene.
  The first distance of 0.28 nm (orange bar) that has a population exceeding 0.3 (red dotted line) was chosen as the effective thickness of the membrane.
  \protect\subref{transfer_diagram} Diagram for identifying transfer events. The green arrows indicate a valid forward transfer event; the grey arrows are not counted.}
  \label{thickness_transfer}
\end{figure}

To gain further insights on the water and ion transfer process, we developed a trajectory-analyzing tool called the ``buffered-region method''.
Our analysis method first estimates the z-coordinate and thickness of the nanoporous graphene sheet.
The z-coordinate is computed as the averaged geometric center of C atoms, and the thickness is estimated as the distance where the radial distribution function (RDF) between pore and water atoms reaches the chosen threshold of 0.3.
As shown in Figure~\ref{thickness}, the membrane thickness turns out to be 0.28 nm from center to surface, though this is rather insensitive to the chosen RDF threshold.
The simulation box is divided into three sections: feed, middle, and permeate, and the events for each individual water molecule crossing the boundary of two sections are recorded.
As illustrated in Figure~\ref{transfer_diagram}, a forward transfer event is counted when a water molecule crosses from the feed to middle region, followed by a crossing from the middle to permeate region; the reverse ordering is used to count backward transfer events. Any other crossing sequences, such as feed to middle then back to feed, are ignored.
The \textit{total exchange rate} is defined as the sum of forward and backward transfer rates, while the \textit{net transfer rate} is the difference between forward and backward transfer rates.
The analysis tool also allows us to access the characteristics of individual transfer events including transfer times and changes in the hydrogen bonding environment.

\begin{figure}[H]
  \includegraphics[width=.6\linewidth]{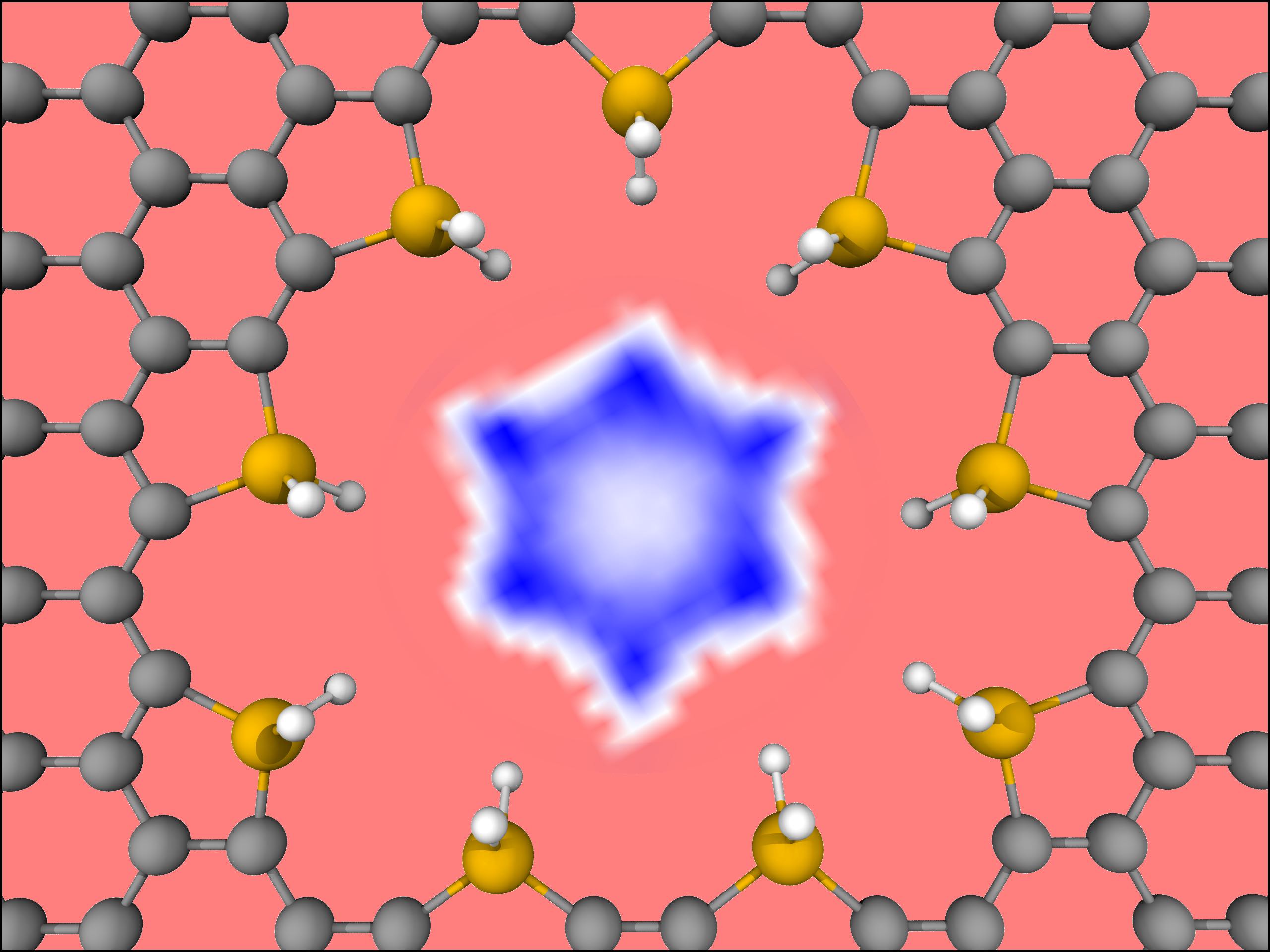}
  \caption{Density map for water molecules inside hydrogenated Si-9 pore. The positions of water molecules are represented by the coordinates of the oxygen atoms. Red color indicates zero density. White area denotes non-zero water density, which is used as the accessible pore area. Darker blue color indicates higher density. Density maps for other pores are shown in the SI section~S7.}
  \label{si9_den_map}
\end{figure}

Our analysis of the water transfer and ion rejection mechanism necessitate two complementary definitions of pore area.
Based on the total water exchange, we calculate the \textit{effective pore area} as:
\begin{equation}
  \label{effective_area}
  A_{\mathrm{pore;eff}} = \frac{ N_{\mathrm{pore}} }{ N_0 } \times A_0,
\end{equation}
where $A_{\mathrm{pore;eff}}$ is proportional to the total water exchange $N_{\mathrm{pore}}$, and $A_0/N_0$ is a normalization factor calculated from water exchange across an unimpeded cross section of the bulk solution.
We also define the \textit{accessible pore area} by integrating the region inside the pore with a nonzero water density; (Figure~\ref{si9_den_map}) 
in contrast to the former definition, the accessible pore area does not depend on dynamical properties such as the total exchange rate.
In practice, the accessible pore area is converged to within 1\% after collecting 100 ns of the simulation data.



All molecular dynamics simulations were performed using the OpenMM package\cite{Eastman:2017kn}. The three-dimensional rendering were generated using VMD.\cite{Humphrey:1996to} Trajectory analyses were carried out in Python using the MDTraj library.\cite{McGibbon:2015fv}

\newpage
\section{Results and Discussion}

\subsection{Optimized Force Field Parameters}
\begin{figure}[H]
    \centering
    \includegraphics[width=0.8\textwidth]{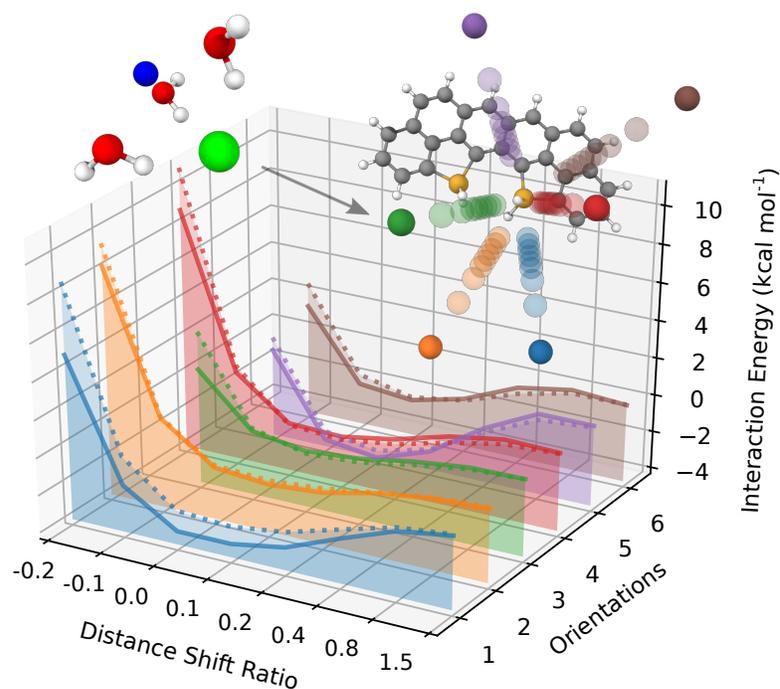}
    \caption{Fitting of the Si--Cl target in various orientations and distances. QM data are plotted in solid lines and MM data are in dotted lines. QM interaction energies larger than 10 \kcal were not fitted. The molecular geometry on the top right shows Cl$^-$ approaching from six orientations, using six different colors corresponding to plot colors. Each colored ball (solid or transparent) represents a Cl$^-$ ion in the [Cl(H$_2$O)$_3$Na] cluster model; the full cluster used in the calculations is shown on the top left. The Na--Cl--Si angles were constrained to be 180$\degree$ in all geometries. Particles: Yellow--Si, Grey--C, White--H, Red--O, Green--Cl$^-$, Blue--Na$^+$.}
    \label{3d_fitting}
\end{figure}

The final force field with optimized parameters accurately reproduces the QM interaction energy data.
For example, Figure~\ref{3d_fitting} shows the SiH--Cl$^-$ interaction energies computed by the quantum mechanical (QM) method compared to molecular mechanics (MM); when the [Cl(H$_2$O)$_3$Na] group approaches the SiH in six different orientations at eight distances, the optimized force field reproduces the quantum interaction energies usually within 1 \kcal.
The set of optimized force field parameters are presented in Supporting Information Section 1.
By comparing the initial and final parameters (SI section 1.2), we confirmed a minimal change for the bonded parameters, while the variation of non-bonded parameters are larger to accommodate water--pore and ion--pore interactions at various geometries.
The complete plot of all \abinitio and fitted MM energy profiles are given in SI section 4.

\begin{figure}[H]
  \subfloat[][]{
    \centering
    \includegraphics[width=.49\linewidth]{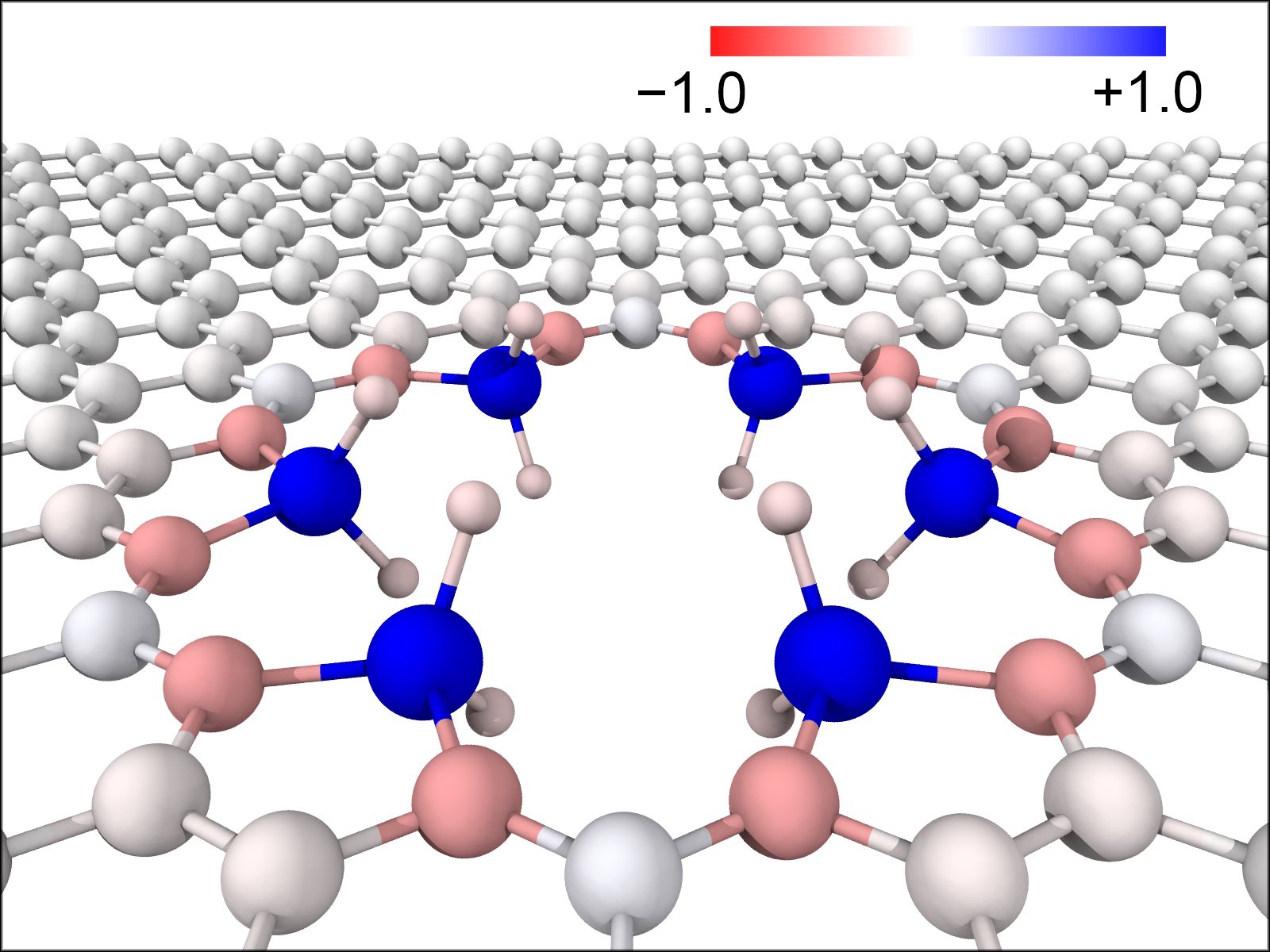}
    \label{subfig:charge_sih}
  }
  \subfloat[][]{
    \centering
    \includegraphics[width=.49\linewidth]{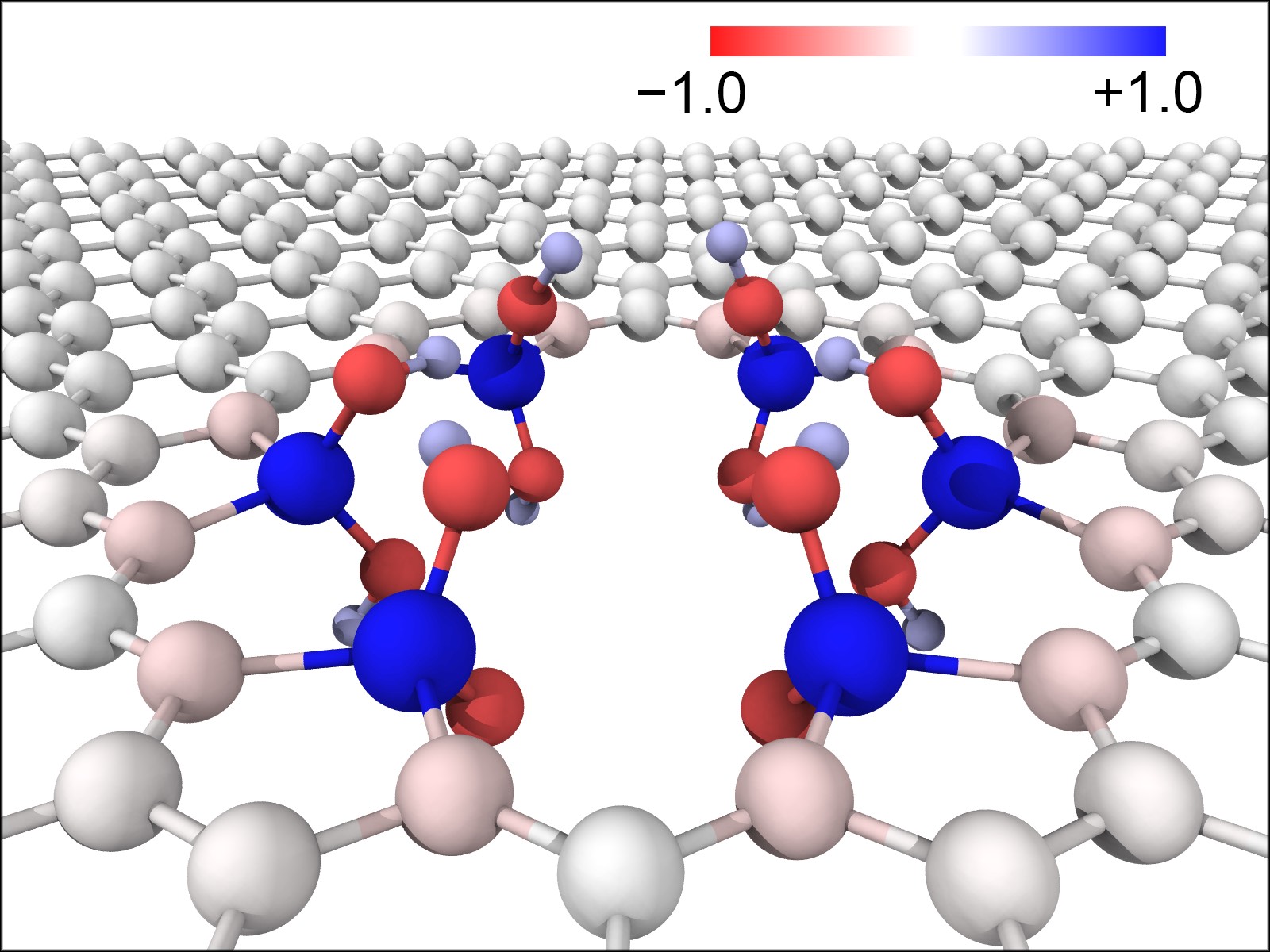}
    \label{subfig:charge_sioh}
  }

  \subfloat[][]{
    \centering
    \includegraphics[width=.49\linewidth]{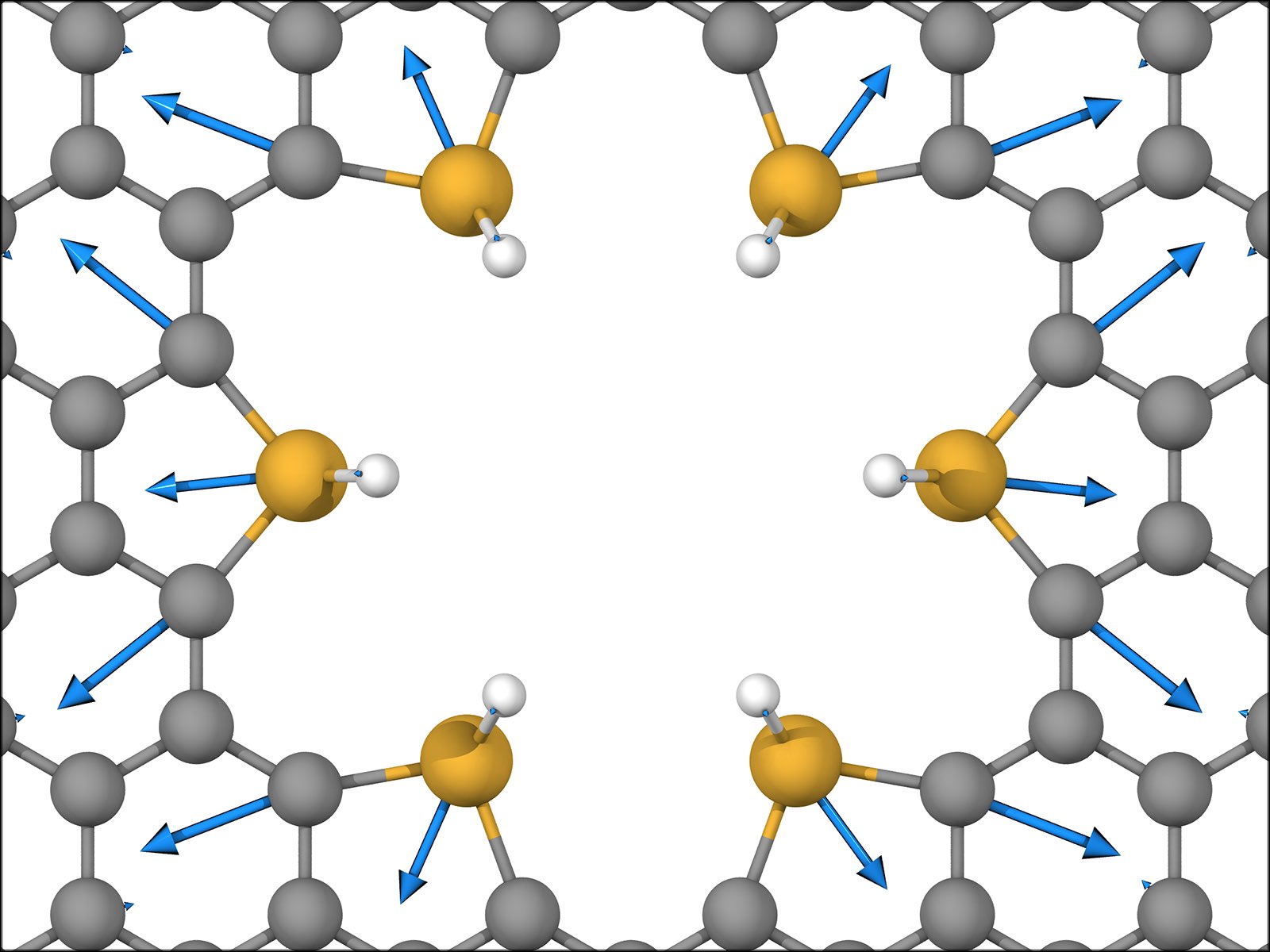}
    \label{subfig:dipole_sih}
  }
  \subfloat[][]{
    \centering
    \includegraphics[width=.49\linewidth]{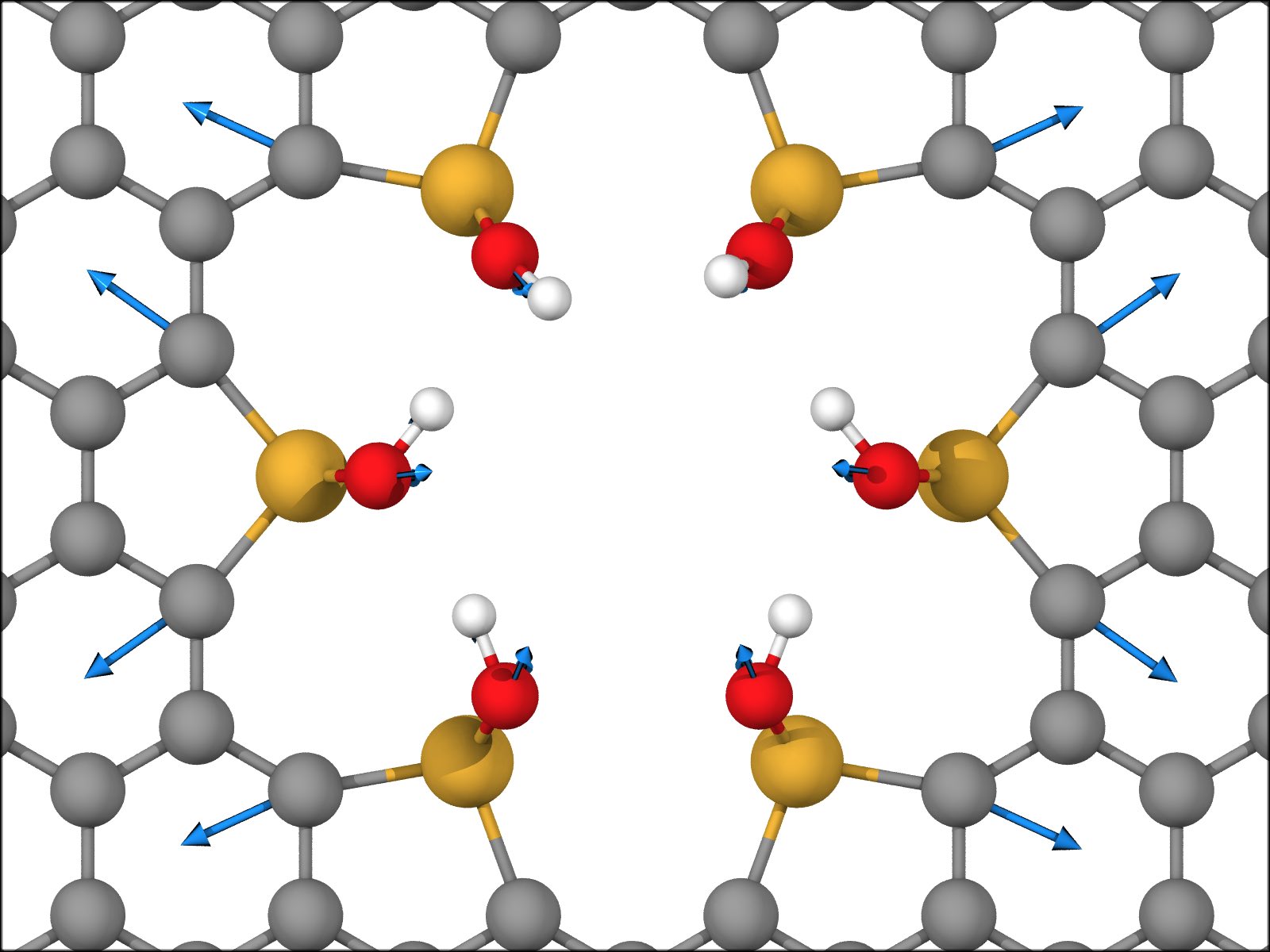}
    \label{subfig:dipole_sioh}
  }
  \caption{Parameterized charge and dipole paramters for the Si-6 pores. \protect\subref{subfig:charge_sih} Si--H charge; \protect\subref{subfig:charge_sioh} Si--OH charge; \protect\subref{subfig:dipole_sih} Si--H dipole; \protect\subref{subfig:dipole_sioh} Si--OH dipole.}
  \label{charge_dipole}
\end{figure}

The partial charges and dipole moments of the optimized pore models are shown in Figure~\ref{charge_dipole}.
For Si--H, the Si atoms on the pore edges carry a positive charge of $+1.03$ , and carbon atoms directly bonded to Si (Cs atom type) carries the most negative partial charge of $-0.33$. (Figure~\ref{subfig:charge_sih})
These Si and C atoms also have the strongest dipole moments, pointing out symmetrically from the pore center. (Figure~\ref{subfig:dipole_sih})
For Si--OH, the edge of the pore is mainly characterized by the negatively charged oxygen atoms, with a partial charge of $-0.76$; (Figure~\ref{subfig:charge_sioh})
the Si atoms bonded to OH groups become more positive ($+1.40$).
The dipole moment on Si atoms in the Si--OH pores are smaller than in their Si--H counterparts (Figure~\ref{subfig:dipole_sioh}).

\subsection{Water Transfer}

\begin{figure}[H]
  \subfloat[][]{
    \centering
    \includegraphics[width=.49\linewidth]{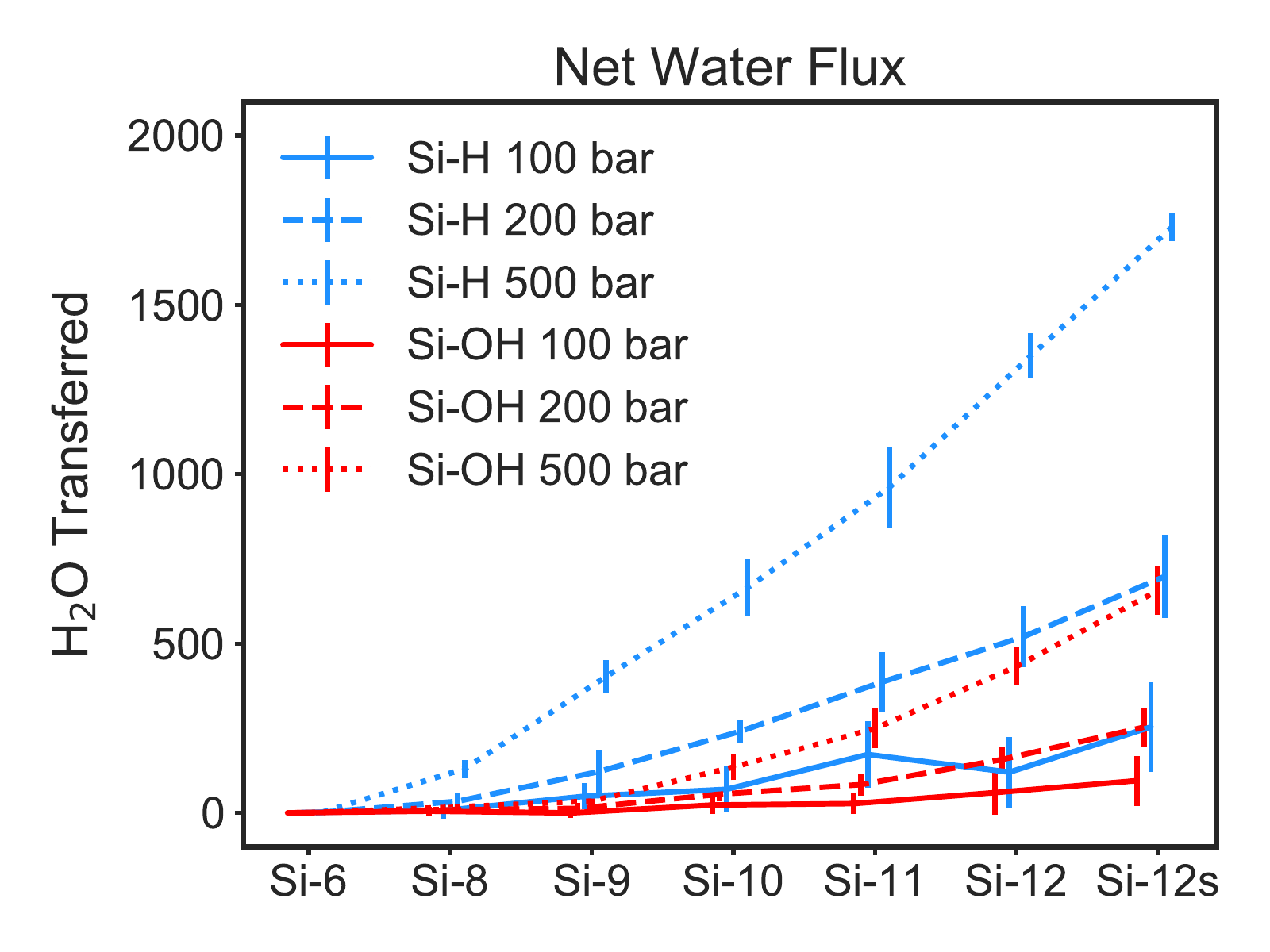}
    \label{subfig:net_water_flux}
  }
  \subfloat[][]{
    \centering
    \includegraphics[width=.49\linewidth]{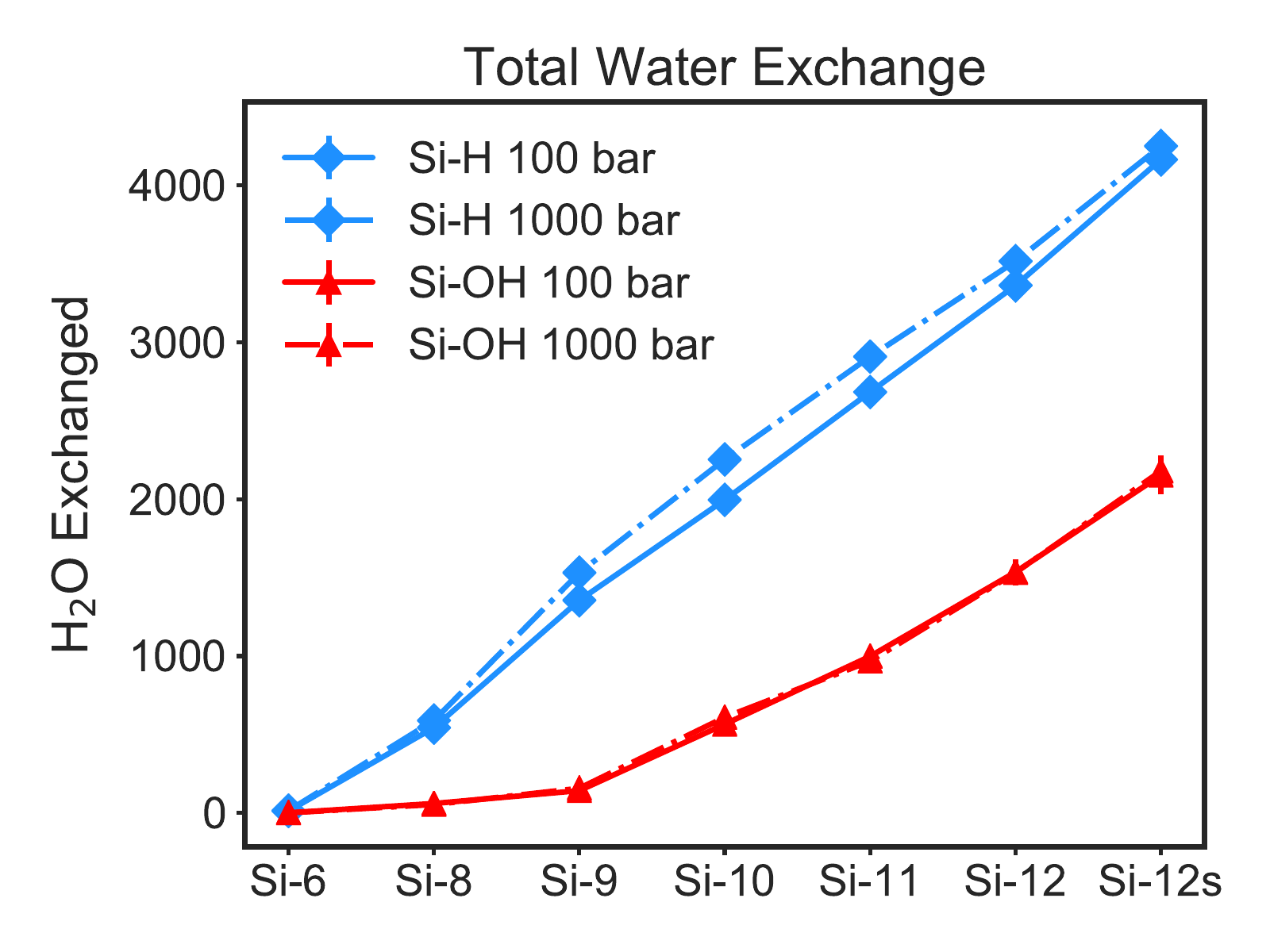}
    \label{subfig:water_exchange}
  }
  \caption{
    \protect\subref{subfig:net_water_flux}
    Number of water molecules transferred (Forward $-$ Backward) in 10 ns. Error bar shows one standard deviation.
    \protect\subref{subfig:water_exchange}
    Total number of water molecules exchanged (Forward + Backward) in 10 ns.
    Data for some driving pressures are not shown for simplicity.
    Error bars are smaller than markers in most cases.
  }
  \label{net_water_flux}
\end{figure}

The net water flux as a function of pore size and driving pressure is plotted in Figure~\ref{subfig:net_water_flux};
an increase in net transfer proportional to applied pressure is observed as expected.
The effect of functional groups is significant, as the flux for each size of the Si--OH pore is at most half of the corresponding Si--H pore.

The total water exchange rates for Si--H and Si--OH pores are shown in Figure~\ref{subfig:water_exchange}; these values are barely affected by the driving pressure, therefore are suitable for quantifying the intrinsic permeability of the pores.
For both Si--H and Si--OH pores, the total water exchange increases with the pore size, reaching a maximum of 400 molecules / ns for the symmetric SiH$_2$-12 pore with ~1 nm diameter; the Si--OH pores also have lowered permeability of $\ge 50$\% relative to their Si--H counterparts, mirroring the trends in water flux.
Three hypotheses are proposed to account for the dramatic difference in pore permeability between Si--H and Si--OH:
1) The OH groups are larger in volume than H, reducing the accessible pore areas of Si--OH pores;
2) The water molecules interact more strongly with Si--OH pores than Si--H pores, resulting in a kinetic barrier when moving through the pores;
3) The ions are impeding water transfer in the Si--OH pores more strongly than in the Si--H pores.

\begin{figure}[H]
  \centering
  \includegraphics[width=.49\linewidth]{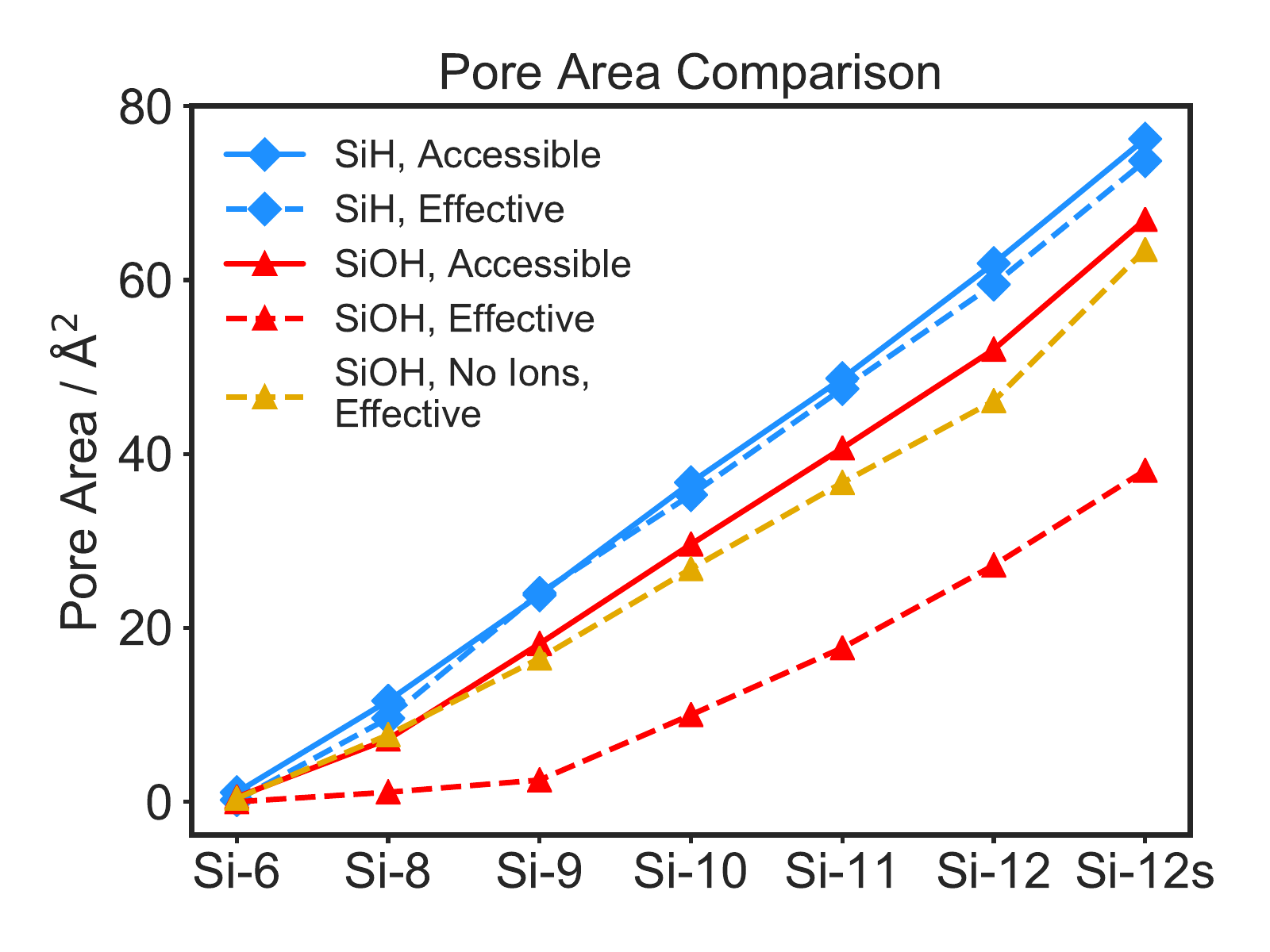}
  \caption{Comparison between accessible pore areas and effective pore areas; the latter is computed using total water exchange under 100 bar driving pressure.}
  \label{pore_area_compare}
\end{figure}

The first hypothesis is tested directly by computing the accessible and effective pore areas.
Figure~\ref{si9_den_map} illustartes how the accessible pore area is computed for the SiH$_2$-9 pore; the water density pattern is hexagonal, indicating that three SiH$_2$ groups interact less closely with water than the other six.
The effective pore area is computed from the total water exchange as in Eq.~\ref{effective_area}, and the comparison of two area measurements is shown in Figure~\ref{pore_area_compare}. 
Despite the fact that water density is not equally distributed inside the pores, we found that the accessible pore area for each Si--H pore matches very well with the effective pore area.
This agreement indicates that pore permeability is proportional to the accessible pore area, consistent with the macroscopic description of maximum flow rate in the limit of short pipe lengths predicted by Bernoulli's principle, $Q_{max}=A\sqrt{2\Delta P/\rho}$. \cite{batchelor2000introduction}
For Si--OH pores, the accessible pore areas are smaller than Si--H by around 5--10 \AA$^2$, but are not enough to rationalize their much smaller effective pore areas; this indicates the larger size of OH functional groups (Hypothesis \#1) cannot fully account for the observed difference between the pores.

After carefully inspecting the simulated trajectories, we found that the Si(OH)$_2$-9 pore is occupied by three or more Na$^+$ ions. (SI section 6)
These ions become immobile as they balance the negative partial charges from oxygen atoms on the Si--OH edge groups, and as a result the water transfer through the pore is greatly slowed down. The average transfer time is 100 ps through the Si(OH)$_2$-9 pore, compared to 10 ps through the SiH$_2$-9 pore.
To confirm the effect of ion blockage, we ran a separate set of simulations with the ions removed.
Without the ions blocking the pore, the water flow rates are much higher, and the resulting effective pore area for Si--OH closely matches the accessible pore area.
This rules out the kinetic barrier hypothesis (\#2) and confirms the ion blockage hypothesis (\#3).


\begin{figure}[H]
  \includegraphics[width=.6\linewidth]{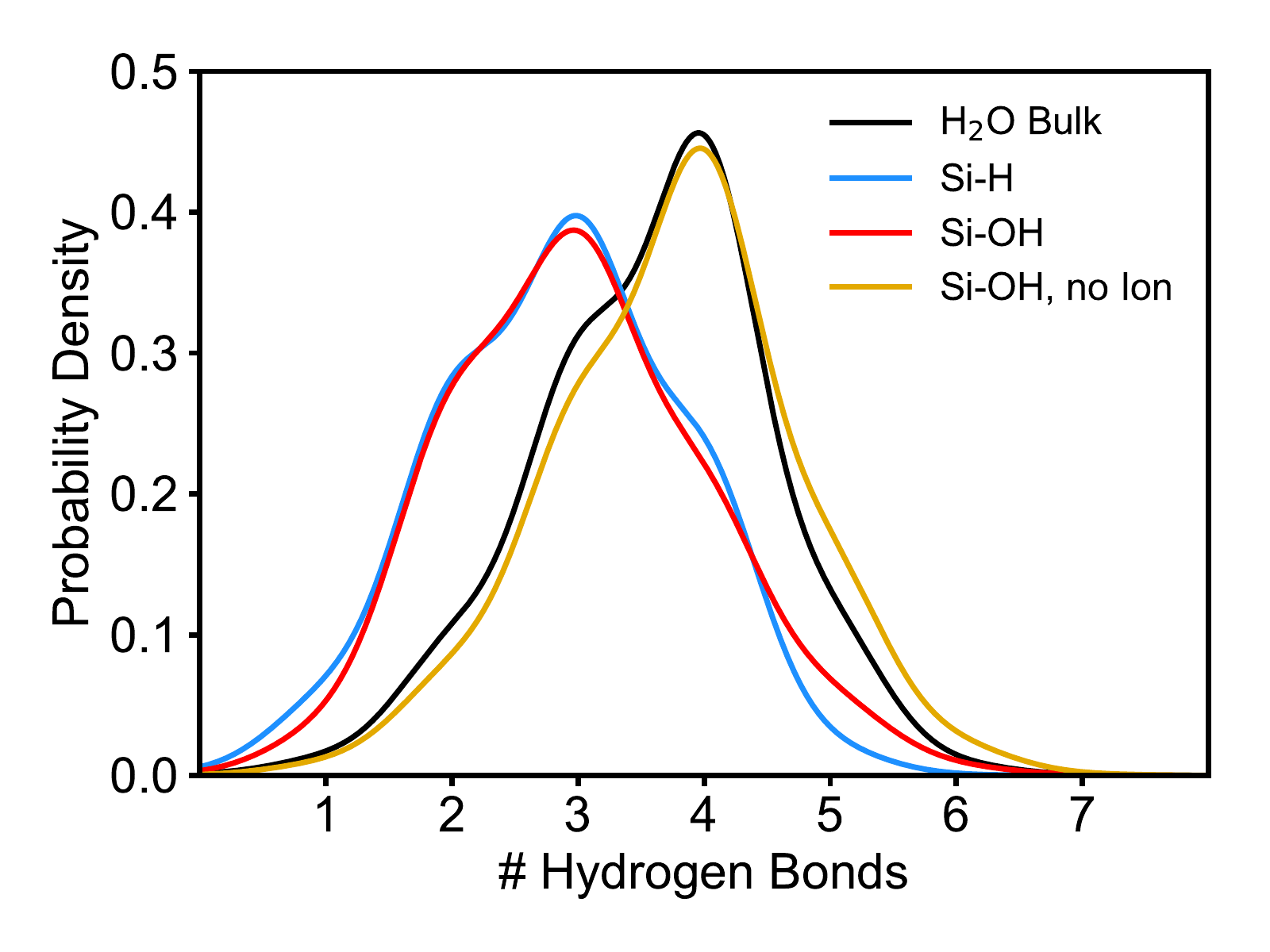}
  \caption{Kernel density estimate (KDE) for number of hydrogen bonds associated with each water molecule. Black: Sampled from bulk water simulation with AMOEBA force field at 348 K; Blue: Water molecules during transfer through the SiH$_2$-9 pore. Red: Water molecules during transfer through the Si(OH)$_2$-9 pore. Yellow: Water molecules during transfer through the Si(OH)$_2$-9 without ion blockage. A Gaussian kernel with bandwidth = 0.5 was used for all plots.}
  \label{hbonds_kde}
\end{figure}

The interaction between water and pore edges can be further understood by quantifying how the hydrogen bonding network changes as a water molecule passes through the pore.
Figure~\ref{hbonds_kde} shows the number of hydrogen bonds ($N_\textrm{hyd}$) each water molecule is involved in, calculated using the hydrogen bond criterion of Baker and Hubbard\cite{Baker:1984cw}.
We computed a reference value of $<N_\textrm{hyd}> = 3.6$ for the average number of hydrogen bonds in bulk water simulated with the AMOEBA force field at 348 K; the KDE shows a clear peak at n=4 reflecting the underlying histogram.
When the analysis is narrowed down to only the molecules undergoing transfer events, the distribution of $N_\textrm{hyd}$ is shifted towards smaller numbers. For the smallest and largest Si--H pore, $<N_\textrm{hyd}> = 1.7\textrm{ and }3.2$ respectively; in particular $<N_\textrm{hyd}> = 2.88$ for SiH$_2$-9.
The reduced hydrogen bonding indicates that water molecules are partially desolvated as they are transferred through the pore, with $<N_\textrm{hyd}>$ decreasing by 0.5--2 depending on pore size.
For the Si(OH)$_2$-9 pore, the OH group pore edges can serve both as hydrogen bond donor and acceptor, and we expected the transferring water to make more hydrogen bonds than in the corresponding SiH$_2$-9 pore; however, the distribution is highly similar to the SiH$_2$-9 pore with $<N_\textrm{hyd}>$ = 2.83.
The reduction in hydrogen bonding is again caused by the Na$^+$ ions filling the pore; when ions were removed from the system, the water molecules transferring through the Si--OH pores have $<N_\textrm{hyd}>$ = 3.76, even a little higher than bulk water.
Additionally, we noticed that the population for zero hydrogen bonds is negligible, suggesting that the evaporation-condensation mechanism\cite{Strong:2016iua} is not playing a significant role in these simulations.

We investigated whether water transfer events are correlated in time by examining the time windows in which a water molecule occupies the middle region for all transfer events.
For the SiH$_2$-9 pore we found that 70\% of water transfer events overlap with other water transfer events in time.
However, 70\% of the entire trajectory contains at least one water transfer event, which means if the water transfer events are distributed randomly with no correlation, there is still a 70\% chance that any one of them would overlap with another.
Therefore, we conclude that existing water transfer events did not have a significant effect on the possibility of the occurrence of another event.

\subsection{Ion Rejection}

\begin{figure}[H]
  \subfloat[][]{
    \centering
    \includegraphics[width=.49\linewidth]{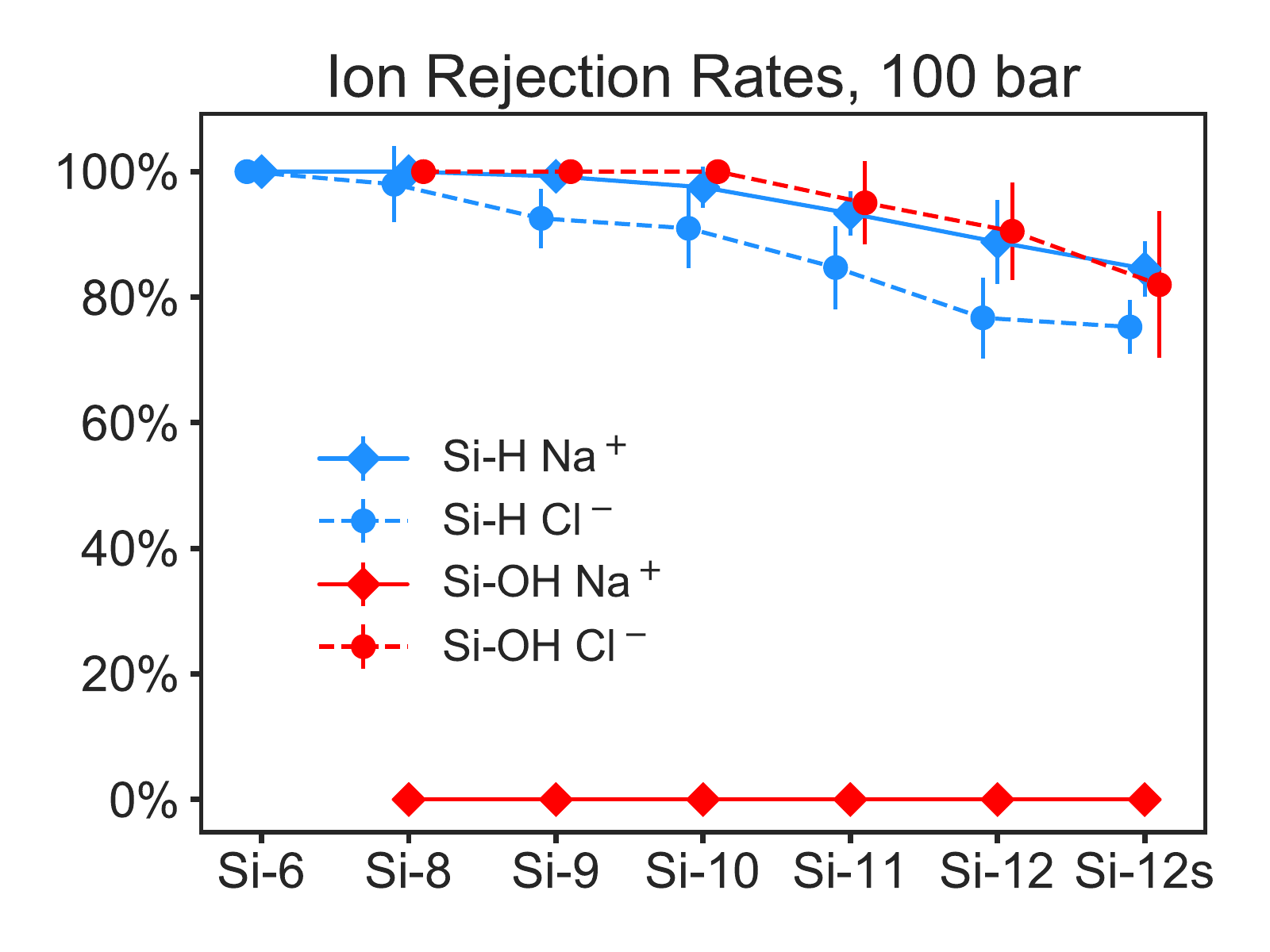}
    \label{subfig:ion_rejection_compare_100}
  }
  \subfloat[][]{
    \centering
    \includegraphics[width=.49\linewidth]{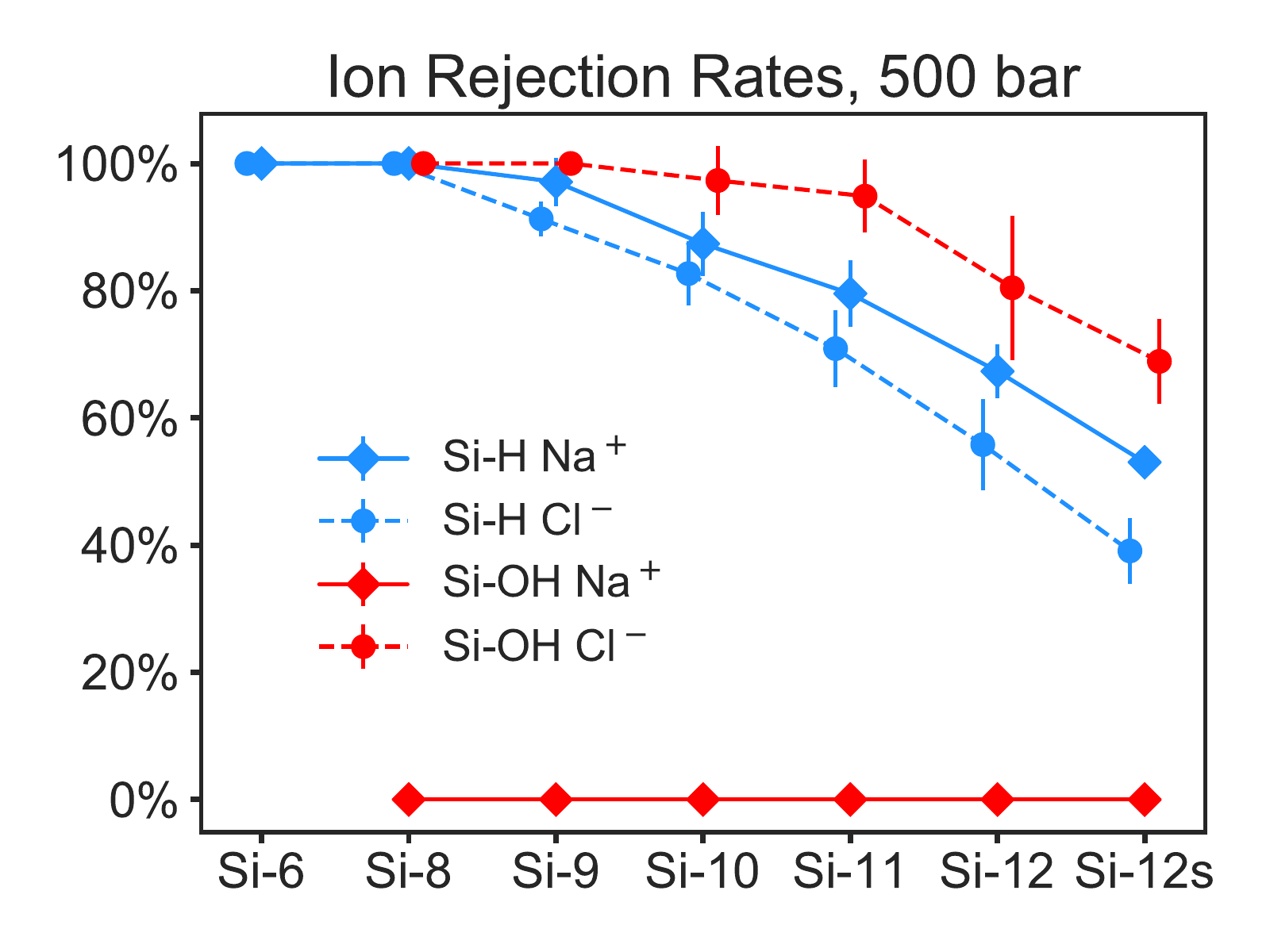}
    \label{subfig:ion_rejection_compare_500}
  }
  \caption{Comparison of the Na$^+$ and Cl$^-$ rejection rates for Si--H and Si--OH pores, under 100 bar driving pressure (left) and 500 bar driving pressure (right). Negative rejection rates were rectified to zero. Extremely small number of water transfer prevents us from calculating a statistically meaningful rejection rate for Si(OH)$_2$-6 pores.}
  \label{ion_rejection_compare}
\end{figure}

The ion rejection rates are calculated by comparing the ion concentration of transferred water molecules and ions relative to the initial feed solution as:
\begin{equation} \label{ion_rejection}
  R\% = 100\% - \left.\frac{\textrm{Transferred Ions}}{\textrm{Transferred Waters}}\right/\frac{\textrm{Initial Ions}}{\textrm{Initial Waters}},
\end{equation}
where $R = 100\%$ corresponds to zero ion leakage, and $R = 0\%$ means the transferred molecules have the same or higher ion concentration than in the feed; only transfer events in the forward direction are counted. 
We find significant differences in ion rejection between positively and negatively charged ions and a strong dependence on the functional groups on the pore edges.
The Si--H pores tend to reject Na$^+$ ions with a higher percentage than the Cl$^-$ ions, as shown in Figure~\ref{ion_rejection_compare};
by contrast, the Si--OH pores selectively blocks the Cl$^-$ ions and does not block Na$^+$ ions at all.
These differences can be explained by two distinct mechanisms of ion rejection and the characteristics of Na$^+$ and Cl$^-$ hydration layers.

\begin{figure}[H]
  \subfloat[][]{
    \centering
    \includegraphics[width=.49\linewidth]{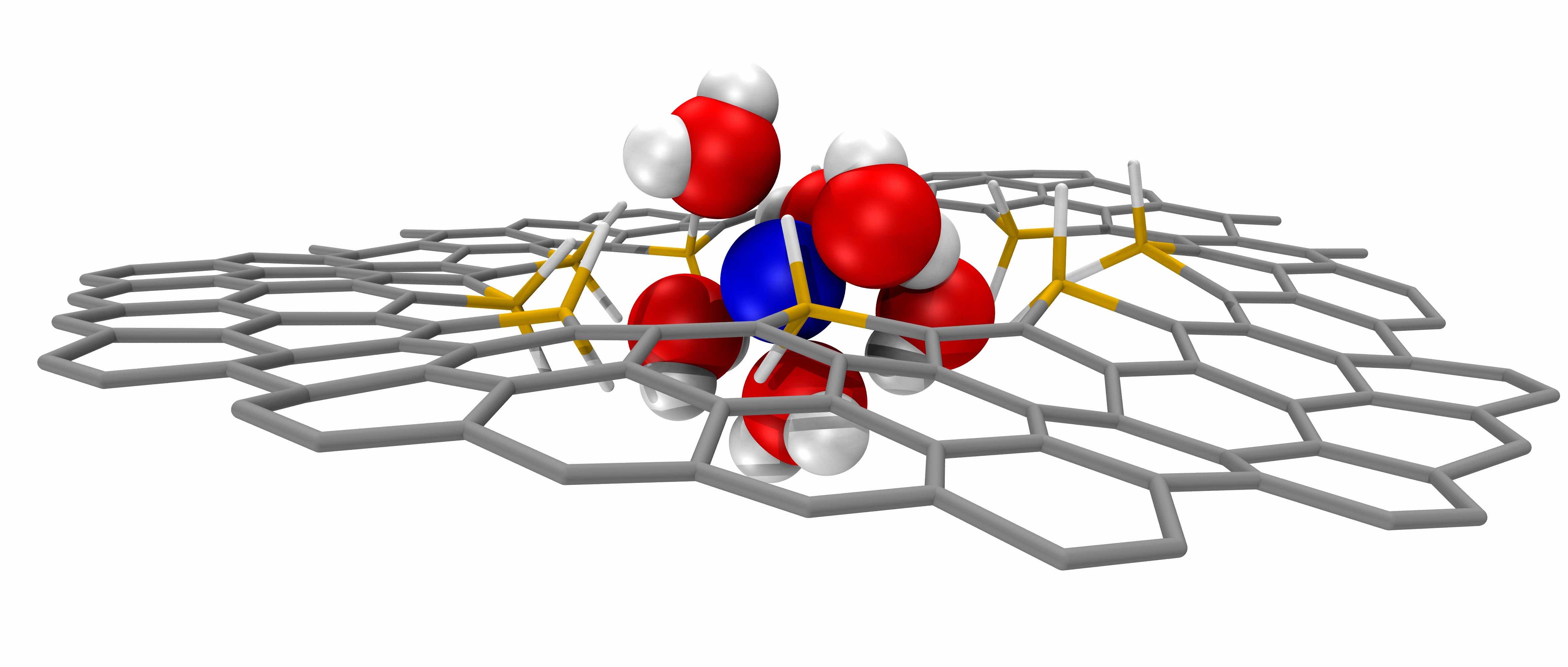}
  }
  \subfloat[][]{
    \centering
    \includegraphics[width=.49\linewidth]{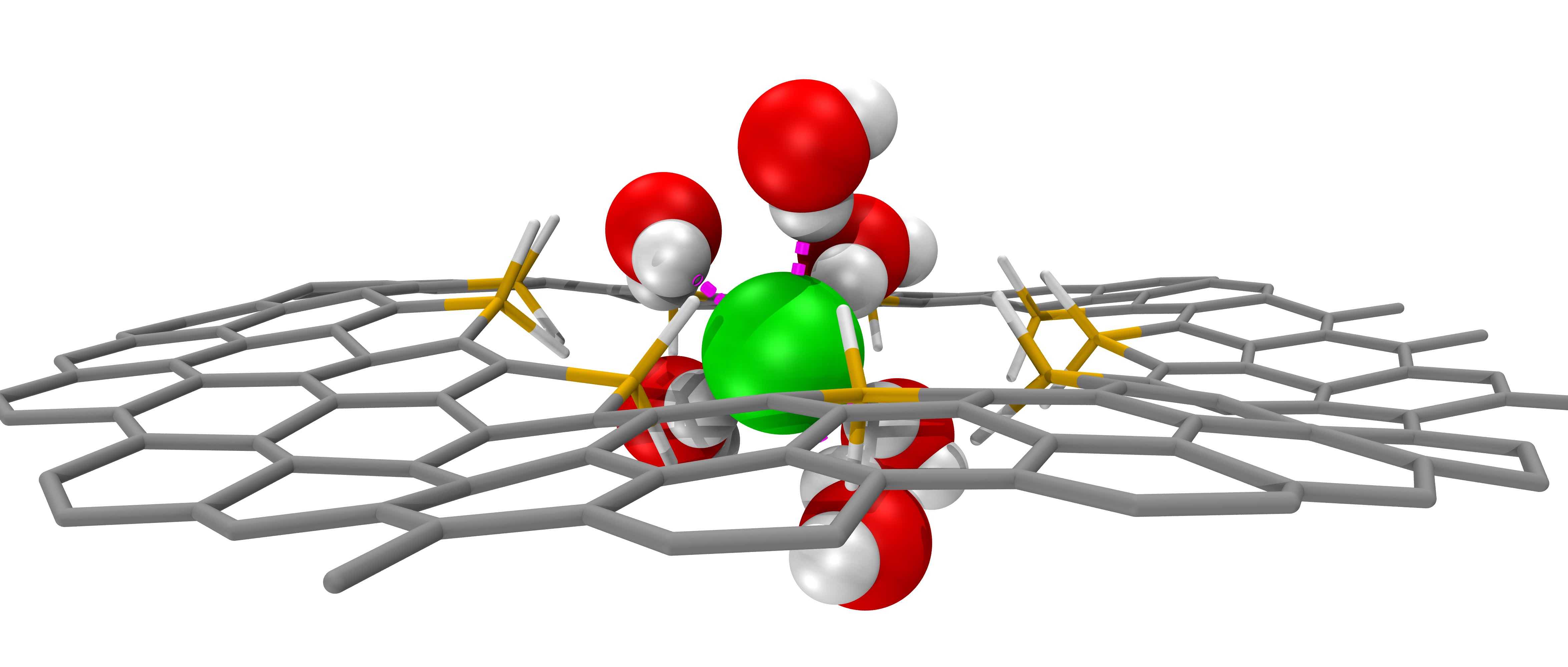}
  }
  \caption{Comparison of Na$^+$ and Cl$^-$ ions with first hydration layer inside a SiH-9 nanopore, indicating the more flexible nature of the Cl$^-$ first hydration shell. This flexibility is postulated to account for the slightly lower ion rejection performance of Si--H nanopores for Cl$^-$ ions.}
  \label{fig:sih_pore_render}
\end{figure}

The Si--H pores do not have strong electrostatic interactions and primarily repel ions via steric hindrance of the ions and their hydration layers, also known as the size rejection mechanism; this is supported by SI Figures~S131--S134 and SI Table~S16 comparing the ion-water RDFs and coordination numbers for ions inside a SiH$_2$-10 nanopore vs. bulk.
The RDF analysis shows that ions permeate the pore with their first hydration shells mostly intact, as indicated by the similarity between the in-pore and bulk RDFs at closer distances than the first trough, as well as the coordination number.
The coordination number decreases by 0.5 for Cl$^-$ ions inside the pore vs. a decrease of only 0.1 for Na$^+$; this indicates the Cl$^-$ hydration layer is more flexible, consistent with previous studies\cite{Impey1983, Mancinelli2007, Laage2007, Bakker2008, Laage2008} and may assist in Cl$^-$ transfer through the pore.
The first trough of the Cl$^-$--O RDF is $>50\%$ lower inside the pore compared to the bulk, and the trough position moves inward by $>0.15$ \AA; this is intuitive because the water O atoms do not directly coordinate to Cl$^-$ and are expected to have greater flexibility in response to a confined environment.
In further support of this picture, the Cl$^-$--H--O angular distribution (SI Figure S135) shows that a broad peak at low values of the acute angle vanishes for ions inside the pore, indicating that water oxygen atoms in the Cl$^-$ hydration layer can undergo significant conformational changes.
Figure~\ref{fig:sih_pore_render} shows the solvation structure of ions found inside a SiH-9 nanopore and provides intuition for how a more flexible hydration layer around Cl$^-$ enables it to permeate the pore more effectively.



\begin{figure}[H]
  \subfloat[][]{
    \centering
    \includegraphics[width=.49\linewidth]{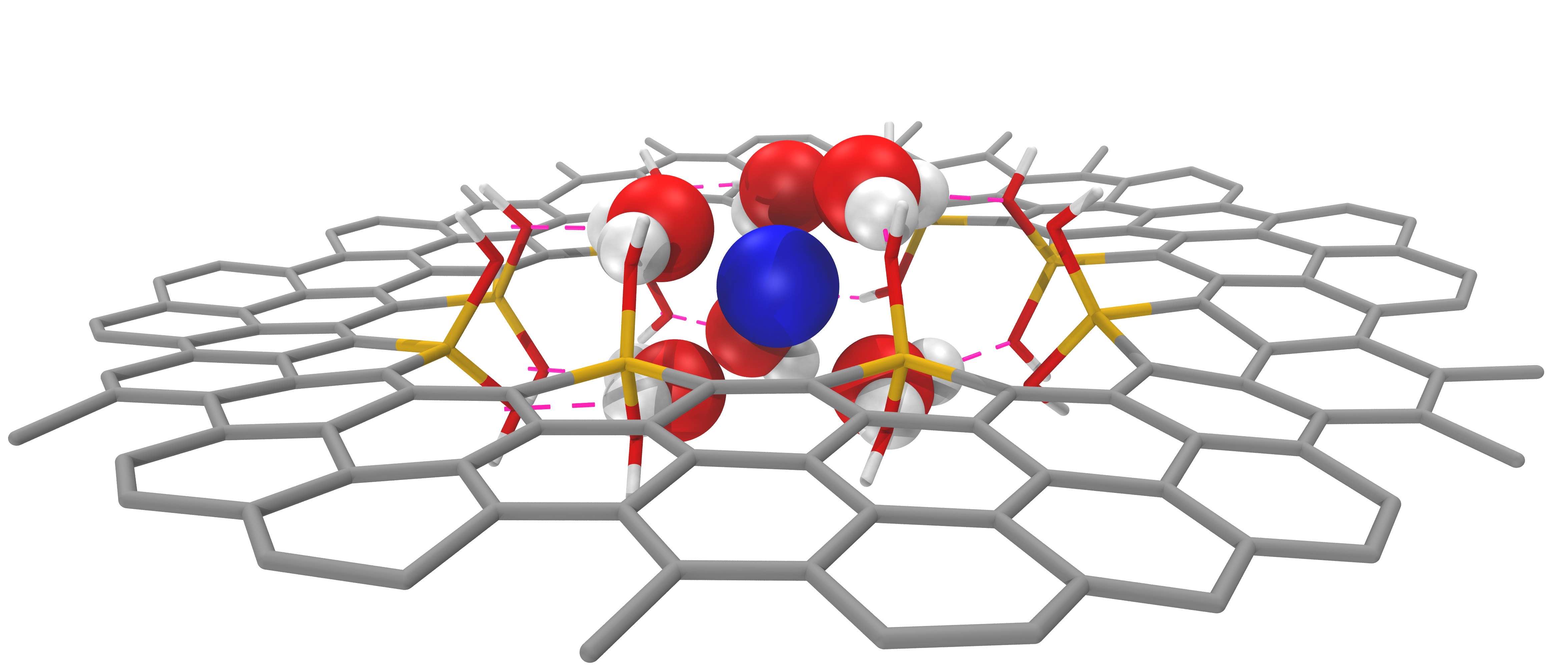}
  }
  \subfloat[][]{
    \centering
    \includegraphics[width=.49\linewidth]{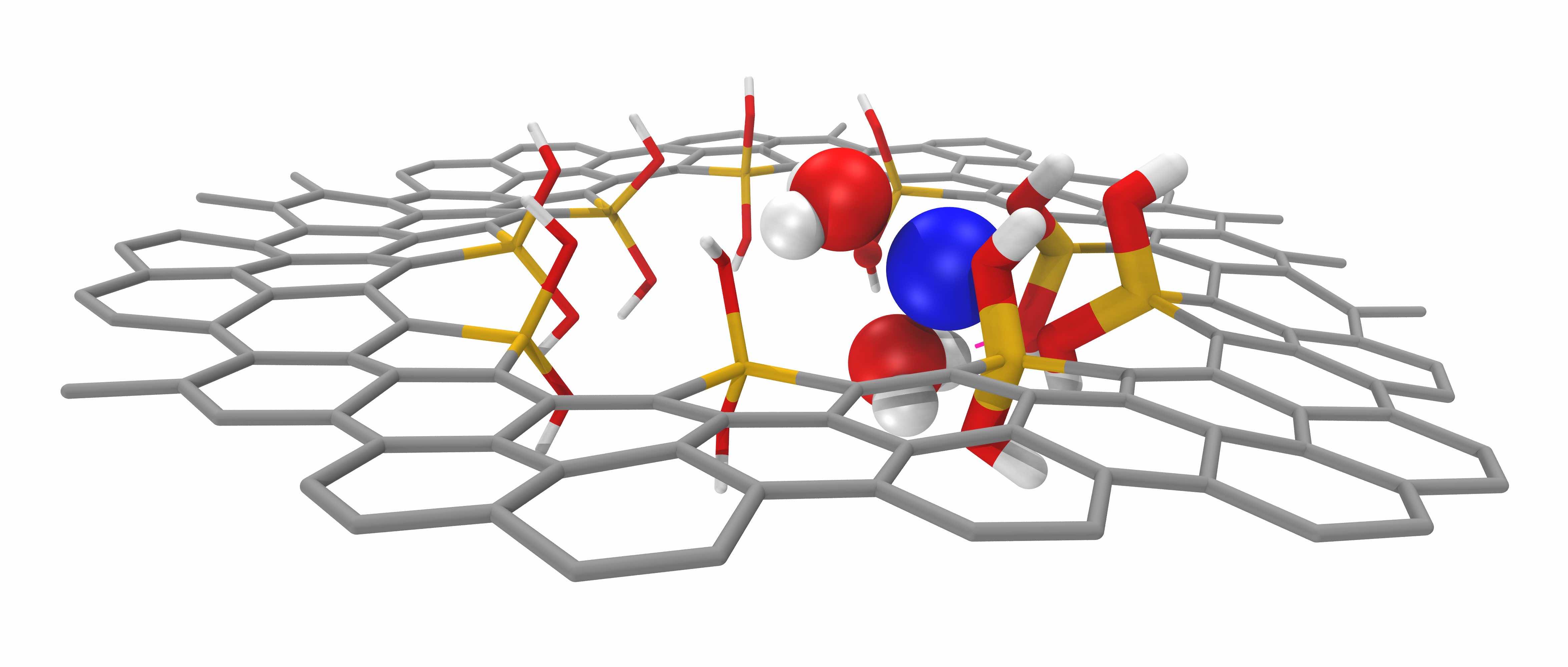}
  }
  \caption{Two snapshots of a Na$^+$ ion blocking a SiOH-9 nanopore. Left: The Na$^+$ hydration layer easily makes hydrogen bonds with the SiOH groups. Right: The SiOH groups can also replace water molecules in the Na$^+$ hydration layer.}
  \label{fig:sioh_pore_render}
\end{figure}

By contrast, the Si--OH pores repel ions primarily by electrostatic interactions (also called the charge rejection mechanism); the OH groups on the pore edge create a negatively electrostatic potential that attracts cations while exclusively blocking anions.
Figure~\ref{fig:sioh_pore_render} shows two snapshots of a Na$^+$ ion occupying the interior of a Si(OH)$_2$-9 nanopore; Na$^+$ may retain its first hydration shell (left panel), or Si--OH groups may take the place of water molecules and coordinate to Na$^+$ directly (right panel). 
This is further substantiated by a large drop in the Na$^+$--O coordination number from 5.7 in the bulk to 4.0 inside the Si--OH pore, which does not occur for Si--H pores (SI Table~S16).
The ability of Si--OH groups to substitute for water in the first hydration shell implies that the accessible pore area for Na$^+$ inside Si--OH pores is greater than in Si--H pores, and explains how Na$^+$ may pass through even the smallest Si(OH)$_2$-6 pore with a diameter less than 6 \AA.
By contrast, when a Cl$^-$ ion permeates the Si--OH pore (which occurs only rarely), we find there are no strong interactions with the pore edges; a snapshot of this system is shown in SI Figure~S136.


Because the permeate solution cannot develop a macroscopic net charge, blocking one type of ion will effectively prevent the large-scale leakage of counter-ions as well; thus we should use the Na$^+$ rejection rates for the Si--H pores, and Cl$^-$ rejection rates for the Si--OH pores to characterize their overall salt rejection performance.
The ion rejection rates for each pore size under different driving pressures are compared in the left and right panel of Figure~\ref{ion_rejection_compare}; the ion rejection rates of Si--H pores are nearly 100\% for small pores and starts to decrease for SiH$_2$-9 and larger pores, in agreement with the size rejection mechanism. 
On the other hand, Si--OH pores have nearly 100\% ion rejection rates even for the relatively large Si(OH)$_2$-11 whose diameter exceeds the hydration shell of Cl$^-$.
These results differ from previous studies\cite{CohenTanugi:2012hs, CohenTanugi:2015fe} in which the authors concluded a lower salt rejection for hydroxylated (C--OH) pores than hydrogenated (C--H) pores; the discrepancy could be attributed to differences in the pore structures used in the simulations, levels of detail in the force fields, and/or simulation analysis approaches.
Ion rejection also decreases with an increase of driving pressure; if we try to maximize the water flux while holding ion rejection at a constant value, we find that smaller pores with higher driving pressures have superior performance to larger pores with lower driving pressure. 
For example, water flux through SiH$_2$-9 under 1000 bar (83 water molecules per ns) is higher than SiH$_2$-11 under 200 bar (39 water molecules per ns), yet both have ion rejection rates above 90\%.


\subsection{Discussions}

In our simulations, the largest Si--H and Si--OH pores with close to 100\% ion rejection rates are SiH$_2$-9 and Si(OH)$_2$-11.
Although Si(OH)$_2$--11 has a larger accessible pore area, its average net water transfer rate (8.5 water molecules per ns) is smaller than SiH$_2$-9 (12.2 water molecules per ns) under 200 bar of driving pressure; the second value is consistent with a previous study that estimates 10.9 water molecules per ns through C--H pore with 5.5 \AA\ diameter.\cite{CohenTanugi:2014jf}
Although the Si--OH pores suffer from cations entering and blocking the pore, the potential transfer rate can be as high as 32.9 water molecules per ns if the ion blockage could be avoided.
In experiments, ion blockage was suspected as the main reason causing the degrading of synthesized nanoporous graphene, but smaller ion concentrations were used (6 mM/mol) and the performance degradation happens on the time scale of 24 hours.
Another possible route to reduce ion blockage is to reduce the negative charge density by replacing a fraction of OH groups with H or other less charged functional groups; ideally a good balance could be achieved that does not diminish the good Cl$^-$ rejection performance.

In the experimental study of Ref.~\cite{Surwade:2015cy}, the authors concluded on a massive 3000 water molecules ns$^{-1}$\ pore$^{-1}$, more than two orders of magnitude higher than our simulated results.
The large difference might be explained by the large area of nanoporous graphene in direct contact with water.
In the experimental setup, the nanoporous graphene prepared by oxygen plasma was suspended over a SiN wafer, in which a 5 $\mu$m diameter hole was made for water to pass through, and the hole area was considered as the total nanoporous graphene area that was operational in the desalination experiment.
However, the area of the nanoporous graphene in direct contact with water is hundreds of times larger than the area of the 5 $\mu$m diameter hole in the SiN wafer.
Therefore, if water could be transported laterally between the graphene and SiN wafer, the nanoporous graphene area responsible for the water transfer may be larger than the hole area.
This hypothesis can be tested by expanding the 5 $\mu$m hole to 10 $\mu$m and even larger, and see if the water transfer rate approaches a limit instead of increasing proportionally to the hole area.

The same paper reports a separate experiment where the authors calculate the flux of forward osmosis (water entering the feed side) to be 70 g m$^{-2}$ s$^{-1}$ atm$^{-1}$.\cite{Surwade:2015cy}
Based on this number and assuming a pore density of 1 per 100 nm$^2$, we calculated the initial water transfer rate under 50 atm of osmotic pressure to be 10 molecules per ns per pore.
This is in close agreement with our simulated forward osmosis transfer rate with 1 mol/L NaCl solution and no driving pressure, which is 6 water molecules per ns per pore for the SiH-9 pore.
The level of agreement between forward osmosis simulations and experiment suggests that the microscopic description of the pore and ion/water transfer is reasonable, and the remarkable desalination performance may originate from macroscopic features of the experimental configuration.

Another possibility for explaining the higher water transfer numbers in experiment is that our optimal pore size could be too small.
We estimated that the optimal pores for high flux and salt rejection should have a diameter of about 6 \AA, which is consistent with earlier theoretical work\cite{CohenTanugi:2012hs} using non-polarizable force fields that predicted high flux and salt rejection with pores approximately 5.5 \AA\ in diameter).
It is possible that long-range charge transfer in the graphene layer could allow nanopores to accumulate significant amounts of net charge, thereby allowing more ions to be rejected via the charge rejection mechanism (though Ref.~\cite{Li:2018eu} predicts an inverse effect when charges are too high).



\section{Conclusions}

In the present work, we investigated the molecular mechanism of water desalination using nanoporous graphene passivated by Si--H and Si--OH groups.
Our simulation used a newly developed AMOEBA-type polarizable force field that reproduces high-level \abinitio quantum chemistry data with good accuracy.
Reverse osmosis simulations were carried out with various pore sizes and driving pressures to measure net water transfer and ion rejection rates;
the effective pore area of Si--H pores based on water transfer rates is commensurate with the solvent-accessible pore area, 
while for Si--OH pores the water transfer is impeded due to Na$^+$ ions occupying the pore interior.
Size-based and charge-based ion rejection mechanisms were discussed for the Si--H and Si--OH types of pores respectively.
Our simulations and analyses revealed complex relationships between water, ions, and pore edges at the nanoporous graphene interface, suggesting that 
intermediate values of pore-ion interaction strengths may provide a route toward optimizing this system.
While our models provide an accurate and physically realistic description of intermolecular interactions between pore edges and water/ions,
they do not account for long-range charge transfer and local deformations of the graphene monolayer that may also impact desalination performance.
Inclusion of these additional effects would further improve on the modeling of nanoporous graphene and further advance our understanding of this promising system.



\begin{acknowledgement}
 This work was supported by a Restricted Gift from Walt Disney Imagineering and ACS-PRF Award \#DNI-58158.
\end{acknowledgement}


\begin{suppinfo}
Complete set of optimized force field parameters, benchmarks on interaction energies, details on the coronene model for fitting graphene parameters, QM vs. MM fitting results, description of the virtual blocking force in the simulation setups, simulation snapshots, water density maps inside pores, full set of simulation data plots, details for No-Ion simulations, estimation of osmotic pressure, ion--water radial distribution functions are available in the supporting information.
\end{suppinfo}


\bibliography{cite}

\clearpage

\clearpage

\clearpage

\end{document}